\documentclass[10pt, conference, compsocconf]{IEEEtran}
\ifCLASSINFOpdf
\else
\fi

\usepackage{mathptmx} 
\usepackage[noend]{algpseudocode}
\usepackage{algorithm}
\usepackage{fancyhdr}
\usepackage[normalem]{ulem}
\usepackage{xspace}
\usepackage[hyphens]{url}
\usepackage[sort,nocompress]{cite}
\usepackage[final]{microtype}
\usepackage{qcircuit}
\usepackage{subcaption}
\usepackage[font=bf]{caption}
\usepackage{tkz-berge}
\usepackage{listings}
\usetikzlibrary{positioning, fit,calc}
\usepackage{amsmath, amssymb}
\usepackage{mdframed}
\usepackage{booktabs}
\usepackage{multirow}

\usepackage[bookmarks=true,breaklinks=true,letterpaper=true,colorlinks,citecolor=blue,linkcolor=blue,urlcolor=blue]{hyperref}
\usepackage[capitalise]{cleveref}

\renewcommand{\paragraph}[1]{\vspace{.5em}\noindent \textbf{#1.}}
\usepackage{listings}
\usepackage[scaled=0.80]{DejaVuSansMono}
\newcommand{\ours}{\textsc{satmap}\xspace}
\newcommand{\qaoa}{\textsc{qaoa}\xspace}
\renewcommand{\leq}{\leqslant}

\DeclareMathAlphabet{\mathcal}{OMS}{cmsy}{m}{n}

\newcommand{\diff}[1]{{\color{black}#1}}

\newcommand{\phys}{\mathit{Phys}}
\newcommand{\edges}{\mathit{Edges}}

\newcommand{\graph}{G}
\newcommand{\physq}{p}
\newcommand{\circuit}{C}

\newcommand{\logics}{\mathit{Logic}}
\newcommand{\logic}{q}

\newcommand{\qnot}{\textsc{not}\xspace}
\newcommand{\cnot}{\textsc{cnot}\xspace}
\newcommand{\cnots}{\textsc{cnot}\text{\small s}\xspace}
\newcommand{\qmr}{\textsc{qmr}\xspace}
\newcommand{\maxsat}{\textsc{maxsat}\xspace}
\newcommand{\sat}{\textsc{sat}\xspace}
\newcommand{\smt}{\textsc{smt}\xspace}

\newcommand{\map}{M}

\newcommand{\seq}[1]{\langle #1 \rangle}
\newcommand{\SWAP}{\textsc{swap}\xspace}
\newcommand{\SWAPs}{\textsc{swap}\text{\small s}\xspace}

\newcommand{\cost}{\emph{cost}}

\newcommand{\vmap}{\mathsf{map}}

\newcommand{\vswap}{\mathsf{swap}}

\newcommand{\cSeff}{\mathit{effect}}
\newcommand{\cEdge}{\mathit{\emph{unique}}}

\newcommand{\true}{\mathsf{true}}
\newcommand{\false}{\mathsf{false}}

\newcommand{\model}{I}
\newcommand{\form}{\phi}
\newcommand{\hard}{\mathit{Hard}}
\newcommand{\soft}{\mathit{Soft}}
\newcommand{\mqtex}{\textsc{ex-mqt}\xspace}
\newcommand{\olsq}{\textsc{tb-olsq}\xspace}
\newcommand{\tket}{\textsc{tket}\xspace}
\newcommand{\sabre}{\textsc{sabre}\xspace}
\newcommand{\mqth}{\textsc{mqth}\xspace}
\newcommand{\nolocal}{\textsc{nl-satmap}\xspace}
\newcommand{\cyclic}{\textsc{cyc-satmap}\xspace}
\newcommand{\tokyo}{Tokyo\xspace}
\newcommand{\archplus}{Tokyo$+$\xspace}
\newcommand{\archminus}{Tokyo$-$\xspace}
\newtheorem{Theorem}{Theorem}

\newtheorem{example}{Example}

\begin{document}
%
\title{Qubit Mapping and Routing via MaxSAT}


\author{\IEEEauthorblockN{Abtin Molavi, Amanda Xu, Martin Diges, Lauren Pick, Swamit Tannu, Aws Albarghouthi}
\IEEEauthorblockA{University of Wisconsin-Madison, Madison, WI, USA \\
\{amolavi, axu44, mdiges, lpick2, stannu, albarghouthi\}@wisc.edu}
}


%


\maketitle

\begin{abstract}
  Near-term quantum computers will operate in a noisy environment,
  without error correction.
  A critical problem for near-term quantum computing is laying out a logical circuit onto a physical device with limited connectivity between qubits. This is known as the \emph{qubit mapping and routing} (\textsc{qmr}) problem,
  an intractable combinatorial problem.
  It is important to solve \textsc{qmr} as optimally as possible to reduce the amount of added noise, which may render a quantum computation useless.
  In this paper, we present a novel approach for optimally solving the \textsc{qmr}
  problem via a reduction to \emph{maximum satisfiability} (\textsc{maxsat}).
  Additionally, we present two novel relaxation ideas that shrink the size
  of the \textsc{maxsat} constraints by exploiting the structure of a quantum circuit.
  Our thorough empirical evaluation demonstrates (1) the scalability of our approach compared to state-of-the-art optimal \textsc{qmr} techniques (\emph{solves more than 3x benchmarks with 40x speedup}),
      (2) the significant cost reduction compared to state-of-the-art heuristic approaches (\emph{an average of $\sim$5x swap reduction}),
      and (3) the power of our proposed constraint relaxations.
  
  \end{abstract}

\begin{IEEEkeywords}
quantum computing, qubit mapping

\end{IEEEkeywords}

%
\IEEEpeerreviewmaketitle

\section{Introduction}
\newcommand{\nisq}{\textsc{nisq}\xspace}

Quantum computers enable efficient simulation of quantum mechanical phenomena, and therefore open up the door to advances in quantum physics, chemistry, material design, optimization, machine learning, and beyond.
Unfortunately, near-term quantum computers face significant reliability challenges as quantum hardware is highly error-prone: quantum bits (qubits) used for computation are sensitive to environmental noise. 
Furthermore, implementing \emph{quantum error correction}~\cite{nielsen2002quantum} to detect and correct hardware errors requires thousands of physical qubits, and therefore is unlikely to become viable soon. 
In the meantime,  near-term quantum computers with several dozens of qubits are expected to operate in a noisy environment without any error correction using a model of computation called {\em noisy intermediate-scale quantum} (\nisq) computing~\cite{preskill2018quantum}.  

A critical problem in \nisq computing is laying out a logical circuit onto a physical device with limited connectivity between qubits. This is known as the \emph{qubit mapping and routing} (\qmr) problem. 
Specifically, we can only apply two-qubit gates on physically adjacent qubits, so we need to move (\emph{route}) qubits to physically adjacent locations.
Qubit routing is a noisy process that can be detrimental to successful execution. Thus, our goal is to lay out the circuit in such a way that minimizes the required routing.

Solving \qmr optimally is known to be \textsc{np}-hard \cite{cowtan2019qubit}.
Thus, a majority of the proposed techniques have been heuristic in nature,
producing suboptimal results \cite{tan2020optimality}.
A small number of techniques have been proposed for solving \qmr optimally,
mostly by reducing the problem to optimizing an objective function subject to constraints, e.g., \emph{integer linear programming} or \emph{satisfiability modulo theories}~\cite{tan2020optimal,wille2019mapping,murali2019noise}.
While such \emph{constraint-based} approaches produce optimal results with minimum noise, they have not been scalable to larger circuits.

\begin{figure}[t]
    \begin{subfigure}[t]{0.49\linewidth}
      \includegraphics[width=.9\textwidth]{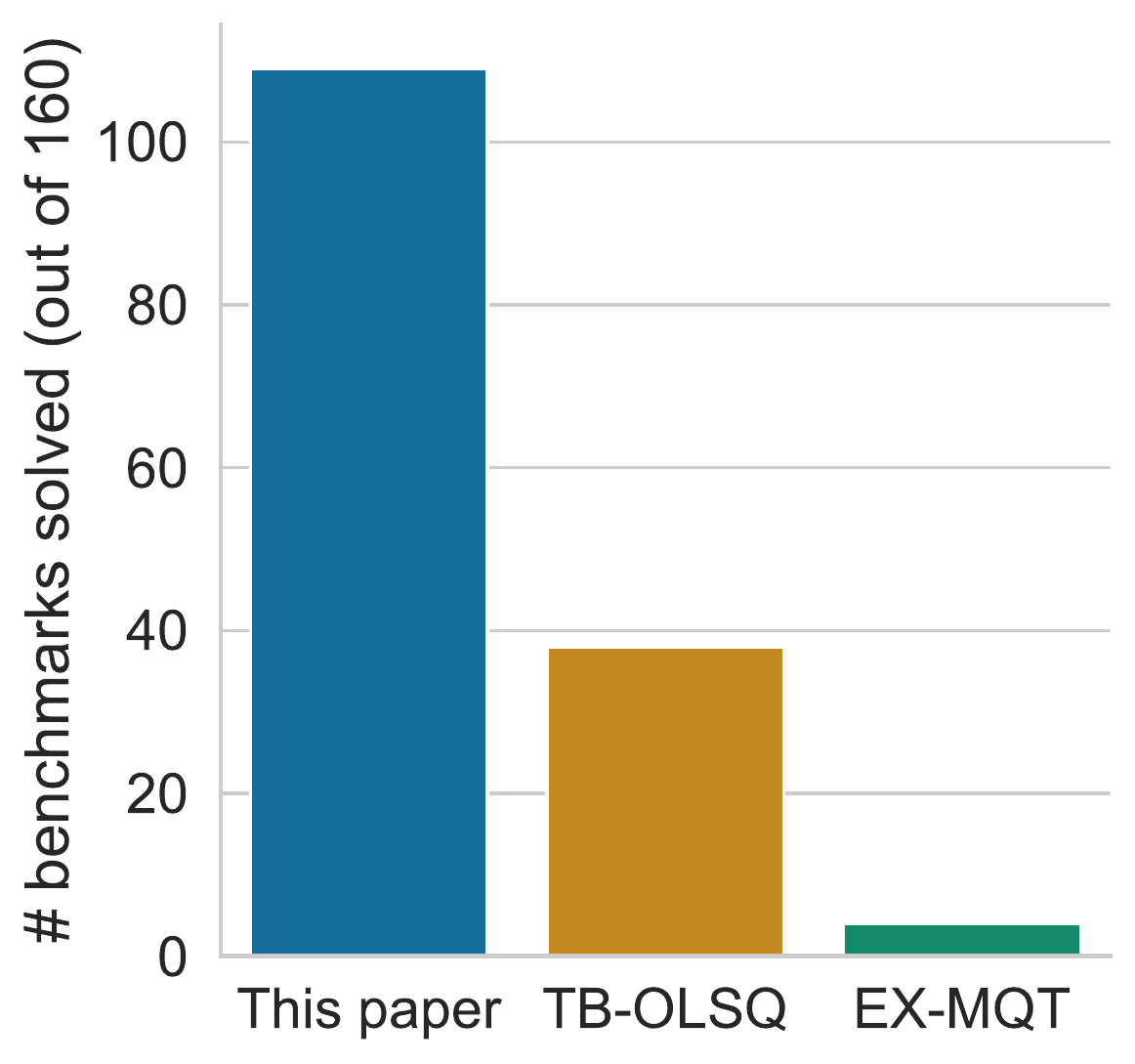}
      \caption{Number of benchmarks solved}
      \label{fig:barchart_scalability}
    \end{subfigure}
    \hfill
    \begin{subfigure}[t]{0.49\linewidth}
      \includegraphics[width=.9\textwidth]{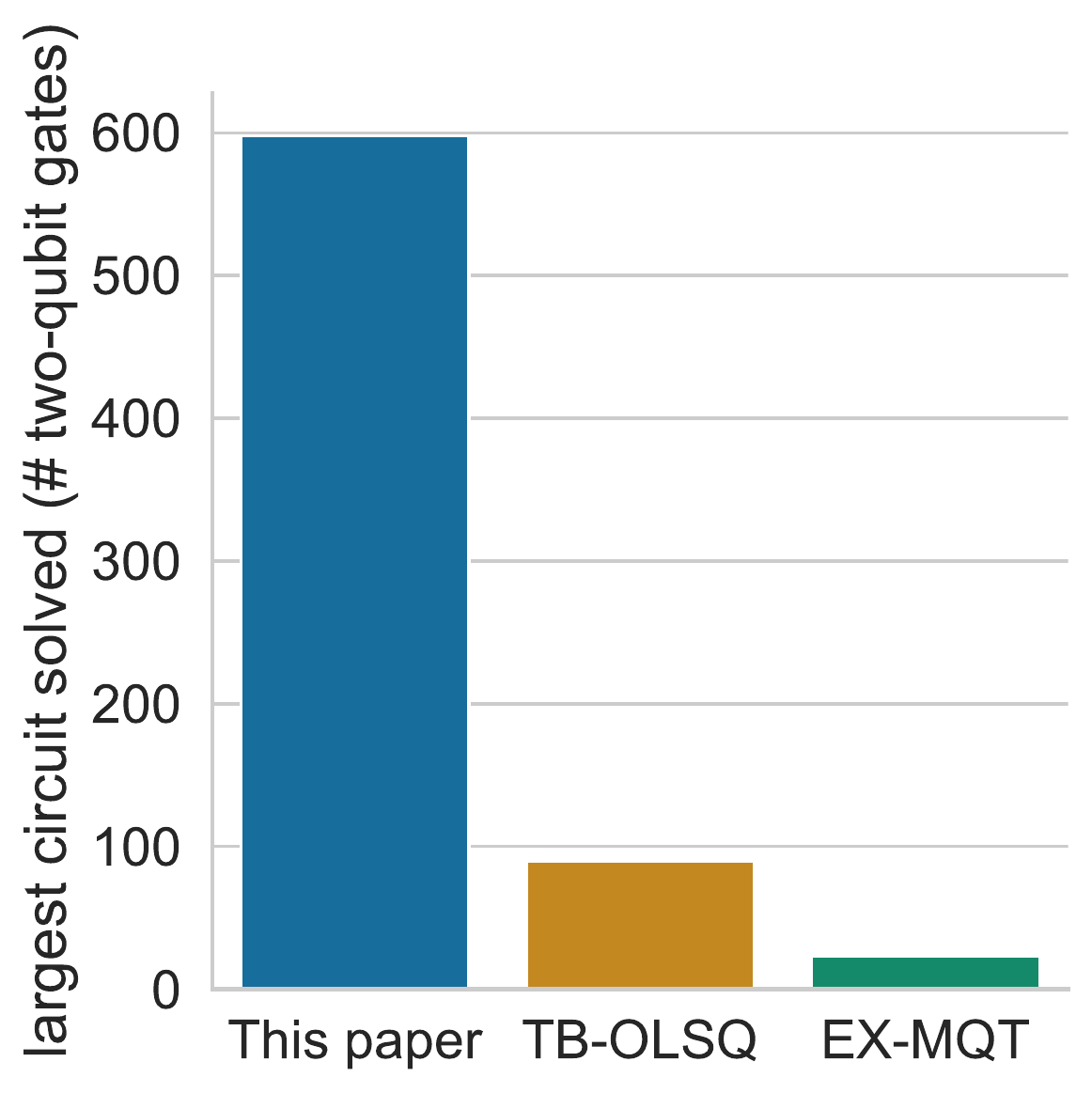}
      \caption{Size of largest circuit solved}
      \label{fig:barchart_scalability_cnots}
    \end{subfigure}
  \caption{Comparison against constraint-based tools}
  \label{fig:comparison_constraint_based}
\end{figure}

In this paper, we propose a novel constraint-based approach that significantly advances the state of the art (see \cref{fig:comparison_constraint_based}).
We believe that scaling constraint-based approaches is an important problem for two reasons: (1) With heuristic \qmr techniques, one can easily add an unacceptable amount of noise for \textsc{nisq} computers, producing uninformative outputs.
(2) Constraint-based techniques present an optimal baseline with which to evaluate the solution quality of heuristic algorithms, and can therefore help us understand and improve their operation.

\paragraph{\qmr as \maxsat}
Our primary insight is that we can reduce the \qmr problem to \emph{maximum satisfiability} (\maxsat)~\cite[Chapter 19]{biere2009handbook}.
\maxsat is the optimization analogue of the Boolean satisfiability (\sat) problem.
While \sat solving is the canonical \textsc{np}-complete problem, the past two decades have witnessed impressive advances in \sat solving with industrial-grade tools applied at scale (e.g., at Amazon~\cite{backes2018semantic}, \sat solvers are invoked millions of times daily).
\maxsat solvers are typically simple loops that repeatedly invoke a \sat solver to get better and better solutions.
Compared to other approaches that use \emph{satisfiability modulo theories} (\textsc{smt}) solvers~\cite{tan2020optimal,wille2019mapping,murali2019noise}, \maxsat solvers are lighter weight as they do not require complex theory-solver interaction.
At a high level, we demonstrate that a \maxsat approach \emph{can} and \emph{should}
be used for solving \qmr constraints.

As summarized in \cref{fig:comparison_constraint_based},
compared to state-of-the-art constraint-based tools~\cite{tan2020optimal,zulehner2018efficient}, our approach can solve significantly more \qmr problems ($\sim$3x) and scale to larger circuits.
In addition, our approach is an \emph{order of magnitude faster} ($\sim$40x) than the fastest constraint-based tool.
Compared to state-of-the-art heuristic-based \qmr tools,
our approach achieves an average of $\sim$3.6x to 7x reduction in the number 
of inserted swap operations.
Further, on $\sim$14\%  of the benchmarks, our approach inserts no swap operations at all.

\paragraph{Sketching-like encoding}
Our \maxsat encoding is inspired by \emph{program sketching}~\cite{solar2008program}, a program-synthesis paradigm where a synthesizer automatically completes \emph{holes} in a program.
In our setting, these \emph{holes} are the routing operations---specifically, \SWAP gates that exchange the contents of two adjacent qubits.
We encode every possible \SWAP as a Boolean variable, where assigning the variable to $\true$ denotes performing a \SWAP of a specific pair of adjacent physical qubits. We therefore ask the \maxsat solver to minimize the number of \SWAP variables set to $\true$.

\begin{figure}[t]
  \begin{subfigure}[t]{0.49\linewidth}
    \includegraphics[width=.9\textwidth]{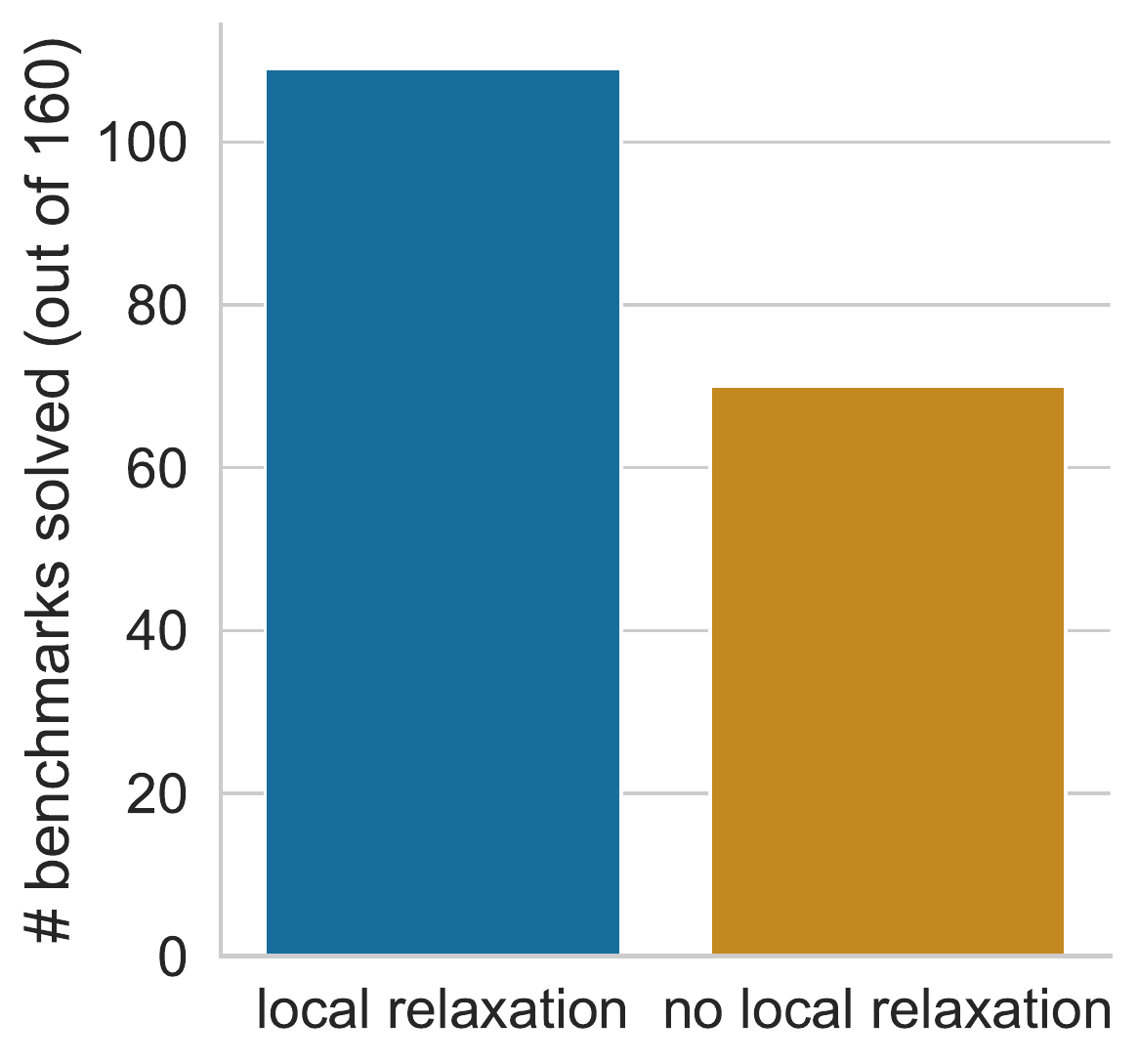}
    \caption{Number of benchmarks solved}
    \label{fig:barchart_scalability_relax}
  \end{subfigure}
  \hfill
  \begin{subfigure}[t]{0.49\linewidth}
    \includegraphics[width=.9\textwidth]{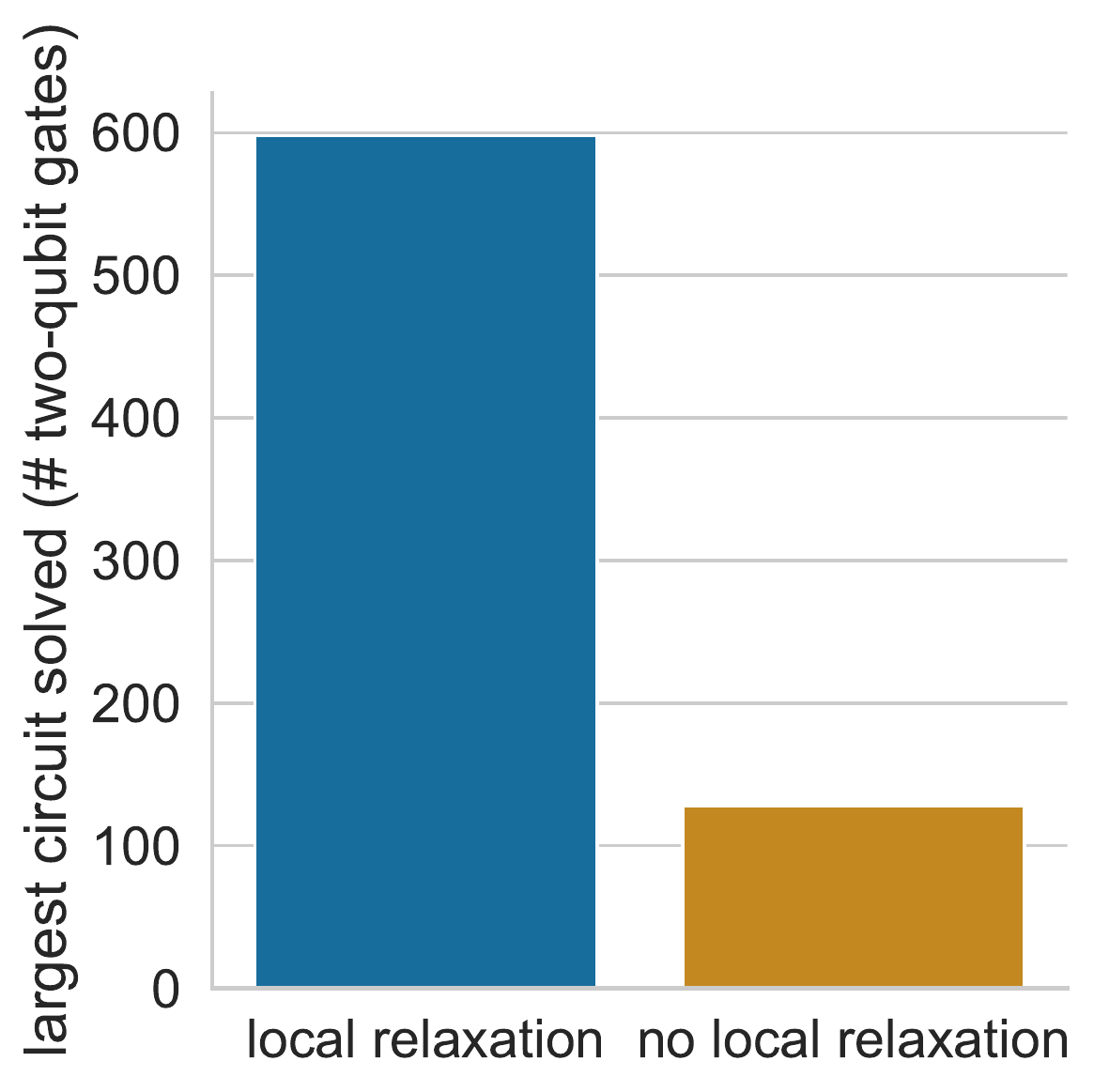}
    \caption{Size of largest circuit solved}
    \label{fig:barchart_scalability_relax_cnots}
  \end{subfigure}
\caption{Comparison against enabling local relaxation}
\label{fig:comparison_slice_sizes}
\end{figure}

\paragraph{Relaxation techniques}
While solving \maxsat constraints results in an optimal \qmr solution,
for large circuits, the \maxsat solver may not be able to efficiently
solve the generated constraints. In this paper, we demonstrate
 a novel relaxation of constraint-based solutions to \qmr, which we call \emph{locally} optimal relaxation.
The idea is to \emph{slice} the circuit horizontally into a number of
consecutive subcircuits and solve a set of smaller \maxsat problems
for each of them.
We demonstrate that our locally optimal relaxation can scale our approach (as shown in \cref{fig:comparison_slice_sizes}) while still producing almost-optimal results (as detailed in \cref{sec:eval}).

Additionally, we present a relaxation for efficiently mapping \emph{cyclic circuits}, like in \emph{quantum approximate optimization algorithms} (\qaoa)~\cite{qaoa}.
Cyclic circuits are where the same subcircuit is repeated more than once.
Instead of generating one big set of constraints for the entire circuit,
we solve a special set of \maxsat constraints only for the repeated subcircuit in isolation,
and then \emph{stitch} the subcircuits back together to generate a mapping for the entire circuit.
Our results demonstrate that this relaxation can make our technique
scale to larger \qaoa circuits.

\paragraph{Contributions} We summarize our contributions as follows:

\begin{itemize}
    \item A novel constraint-based approach for optimally solving the qubit mapping and routing problem via a reduction to maximum satisfiability (\maxsat).
    \item A locally optimal constraint relaxation based on circuit slicing.
    \item A specialized constraint relaxation for cyclic circuits, e.g., as in \qaoa.
    \item A thorough empirical evaluation demonstrating (1) the scalability of our approach compared to state-of-the-art constraint-based techniques,
    (2) the significant cost reduction compared to heuristic approaches,
    and (3) the power of our proposed constraint relaxations.
\end{itemize}

\section{An Illustrative Example}
\label{sec:example}

In this section, we provide background on the qubit mapping and routing (\qmr) problem
and walk through a running example that motivates our approach.

\begin{figure}[t!]
  \centering
  \includegraphics[scale=1.4]{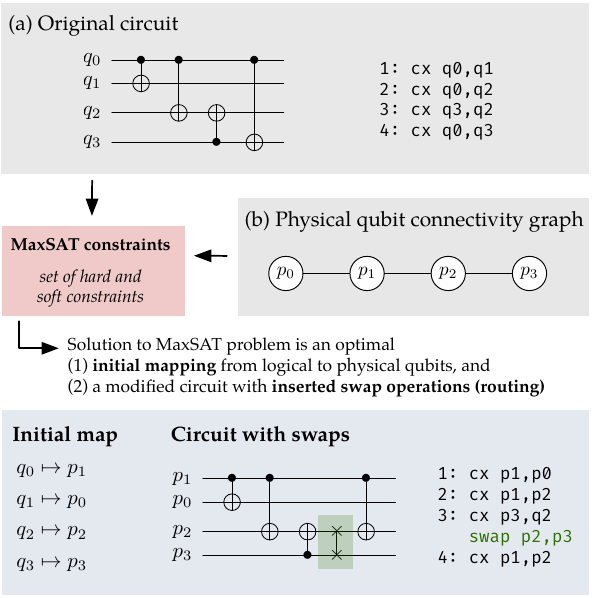}
  \caption{Running example and overview}
  \label{fig:example}
\end{figure}

\paragraph{\qmr primer}
Quantum computers typically support two kinds of operations, single-qubit gates (e.g., \qnot, analogous to a bit flip), and two-qubit gates (e.g., \cnot, analogous to an exclusive or). 
Due to various physical design constraints, \textsc{nisq}-era quantum computers support two-qubit operations only between certain pairs of physical qubits as described by a \emph{connectivity graph}. 
Consider, for instance, the simple connectivity graph in \cref{fig:example}(b),
which illustrates a small device with four physical qubits, $\physq_0,\ldots,\physq_3$.
Edges between physical qubits denote whether we can perform two-qubit operations between them.
For example, we can perform a two-qubit operation over $\physq_0$ and $\physq_1$, but not $\physq_0$ and $\physq_2$.

In order to execute a quantum circuit on a particular device,
the compiler \emph{maps}
the logical qubits that appear in the circuit to appropriate physical qubits such that every two-qubit gate can be applied.
Consider the circuit in \cref{fig:example}(a);
the first gate is a \cnot between logical qubits $\logic_0$ and $\logic_1$
(denoted in the assembly code on the right as \texttt{cx q0,q1}).
Therefore, the logical qubits $\logic_0$  and $\logic_1$  should be mapped
to physical qubits that are adjacent in the connectivity graph, 
e.g., physical qubits $\physq_3$ and $\physq_4$ in the graph in \cref{fig:example}(b).

Typically, a static initial map does not suffice and so the map has to be transformed
during circuit execution to accommodate for other two-qubit gates later on in the circuit.
This process, called \emph{routing}, is achieved by inserting \SWAP operations,
which exchange the values of two connected qubits.
For instance, suppose we want to perform a two-qubit gate on 
$\physq_1$ and $\physq_3$. They are not connected in the connectivity graph;
therefore, we need to bring them \emph{next to each other}.
One way to do so is to swap the qubits $\physq_2$ and $\physq_3$.

Our goal is to solve the \qmr problem \emph{optimally}:
\begin{center}
\emph{Find an initial map that requires the least routing (number of swap operations to be inserted).}
\end{center}

Each additional gate in the circuit
increases the probability of error. In particular, two-qubit gate error rates are significantly
higher than one-qubit gate error rates. In addition, two-qubit gates such \SWAPs have significantly longer gate latency, which can make qubits prone to decoherence errors, so minimizing the number of \SWAPs is critical. 

\paragraph{Our \maxsat approach}
Finding an optimal \qmr solution is 
a combinatorially challenging problem; indeed, it is \textsc{np}-complete.
In this paper, we capitalize on the success of satisfiability (\textsc{sat})
solvers for finding satisfying assignments of Boolean formulas.
E.g., for a Boolean formula $a \land \neg b$, setting $a$ to $\true$ and $b$ to $\false$
is a satisfying assignment.
While the satisfiability problem is the quintessential \textsc{np}-complete problem,
algorithmic and engineering progress in satisfiability has made \textsc{sat} solving practical in many instances~\cite{biere2009handbook}.
Since \qmr is an optimization problem, we use a \maxsat solver, which builds upon a \textsc{sat} solver to find an optimal satisfying assignment.

Roughly speaking, given a circuit and a connectivity graph, we generate a set of Boolean formulas (constraints) whose optimal satisfying assignment corresponds to an initial mapping of logical to physical qubits and a set of \SWAP operations to be inserted before each two-qubit gate.
Specifically, the crucial bit of our encoding is that we model all possible \SWAP operations as Boolean variables;
for example, \texttt{swap p0,p1} would have a corresponding Boolean variable
in every location it could be placed in the circuit.
Then, if a Boolean variable is assigned to $\true$ in the solution
to the \maxsat problem, the corresponding \SWAP is inserted into the circuit; otherwise, it is not.

\cref{fig:example} provides an example of the \qmr problem. The circuit \cref{fig:example}(a) applies 
two-qubit operations between the logical qubit $\logic_0$ and three other logical qubits, but every physical qubit is only connected to at most two other physical qubits.
Therefore, \SWAPs are needed here.
It turns out that inserting a single swap is sufficient for this example. 
\cref{fig:example}(bottom) shows an optimal \qmr solution that can be discovered by solving the \maxsat constraints (the inserted \SWAP is highlighted in green).

A \maxsat solver is typically implemented as a loop that queries a \sat solver
for better and better solutions, until it arrives at an optimal one.
Therefore, a benefit of using a \maxsat solver is that, even for large circuits where the solver cannot efficiently find an optimal solution, the solver may be terminated early to extract the best solution found so far (if it has progressed past the first loop iteration).

\paragraph{Slicing and cyclic circuits}
As mentioned previously,
for large circuits, the \maxsat solver may not be able to efficiently
solve the generated constraints.
In the worst case, the \maxsat solver will not even find a non-optimal solution to the \qmr problem in a feasible amount of time.
In this paper, we demonstrate a locally optimal relaxation of the constraints,
in which we slice the circuit horizontally into a number of
consecutive subcircuits and solve a set of smaller \maxsat problems
for each of them.
This relaxation allows us to scale our approach to larger circuits by sacrificing a guarantee of global optimality.

\begin{figure}[t!]
  \centering
  \includegraphics[scale=1.4]{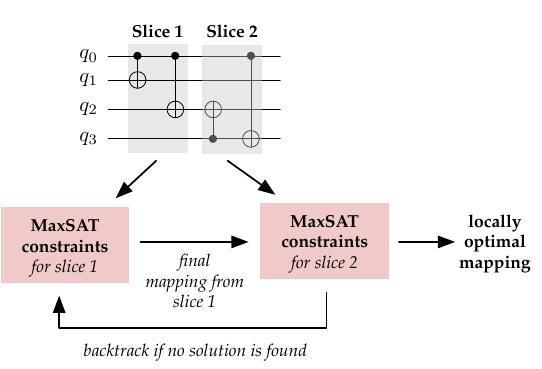}
  \caption{Illustration of our locally optimal relaxation}
  \label{fig:example2}
\end{figure}

This idea is illustrated for our running example in \cref{fig:example2}.
First we slice the circuit into two slices (it could be more, but we stick to two slices for illustration), as shown in the shaded areas.
We solve the \maxsat constraints for the first slice;
this generates an optimal solution for the slice in isolation.
We then take the final mapping from this solution---i.e., the mapping 
at the end of slice 1 after all swaps have been executed---and add it to the
constraints for solving slice 2.
As we describe in \cref{sec:local}, in some cases we need to backtrack
as the solution of slice 1 may be incompatible with slice 2.

We also present an analogous idea for efficiently solving \qmr for \emph{cyclic circuits}, 
which consist of repeated instances of the same subcircuit.
Rather than generating a monolithic set of constraints for the entire circuit at once,
we instead generate and solve a special set of \maxsat constraints only for the repeated subcircuit in isolation.
We then stitch copies of the solution to generate a mapping for the entire circuit.
We describe this idea in \cref{sec:cyclic}.

The two aforementioned relaxations can be easily composed
by slicing subcircuits of a cyclic circuit,
thus exploiting both the cyclic and slice-like structure of a circuit.

\section{Qubit Mapping and Routing}
\label{sec:probstatement}
We now define the qubit mapping and routing problem.

\paragraph{Connectivity graph} We will use $\graph = (\phys, \edges)$ to denote a connectivity graph between physical qubits on a quantum device, where
\begin{itemize}
  \item $\phys = \{\physq_0, \physq_1,\ldots\}$ is the set of physical qubits and
  \item $\edges \subseteq \phys \times \phys$ is the set of edges connecting physical qubits.
\end{itemize}
Graph edges denote on which pairs of physical qubits we can perform two-qubit operations.

\paragraph{Quantum circuit} We will use $\circuit$ to denote a quantum circuit
over logical qubits, $\logics = \{\logic_0, \logic_1,\ldots\}$.
Specifically, a circuit $\circuit$ is a sequence of gate applications, where each gate is an operation that applies to one or two logical qubits.
We will use $g_k$ to denote the $k$th gate in the circuit, $g_k(\logic)$ to denote the application of the one-qubit gate $g_k$ to logical qubit $\logic$, or 
 $g_k(\logic, \logic')$ to denote the application of the two-qubit gate $g_k$ to qubits $\logic$ and $\logic'$.

\paragraph{Qubit map} Given a circuit $\circuit$ and an undirected connectivity graph $\graph$, a \emph{qubit map} $M : \logics \to \phys$ is an injective function from logical qubits  to physical qubits.

Our goal is to find a \emph{map sequence} $\seq{\map_1,\ldots,\map_{|\circuit|}}$, where $|\circuit|$ is the number of gates in the circuit, such that if the $k$th gate in the circuit is a two-qubit gate $g_k(\logic,\logic')$, then $$(\map_k(\logic), \map_k(\logic')) \in \edges,$$ i.e., logical qubits $\logic$ and $\logic'$ are mapped to physical qubits that are connected in the connectivity graph.

\begin{example}
  Recall our running example.
  The initial map, $\map_1$, is shown in  \cref{fig:example}(bottom).
  Observe that logical qubits $\logic_0$ and $\logic_1$ are mapped to adjacent physical qubits, $\physq_1$ and $\physq_0$, i.e., 
  $(\map(\logic_0),\map(\logic_1)) \in \edges$.
  \end{example}

\paragraph{\SWAP operations} We will use $s(\physq, \physq')$ to denote the $\SWAP$ operation that swaps the physical qubits $\physq$ and $\physq'$.

Suppose that in qubit map $\map$ we have $\map(\logic) = \physq$ and $\map(\logic') = \physq'$.
Applying $s(\physq, \physq')$ in $\map$
results in a new map $\map'$ that is just like $\map$ but where $\map'(\logic) = \physq'$ and $\map'(\logic') = \physq$.
Note that $\SWAP$ operations are only allowed on pairs of connected physical qubits, i.e.,  $(\physq, \physq') \in \edges$.

\paragraph{Optimal qubit mapping and routing (\qmr)}
An \emph{optimal} solution to the \qmr problem is a map sequence $\seq{\map_1,\ldots,\map_{|\circuit|}}$ that minimizes the cost of \emph{routing} qubits between adjacent maps in the sequence; formally:
\begin{equation*}\label{eq:opt}
\min \sum_{i=1}^{|\circuit|-1} \cost(\map_i, \map_{i+1})  
\end{equation*}
where $\cost(\map, \map')$ is the smallest number of $\SWAP$ operations needed to go from $\map$ to $\map'$.

\begin{example}
  Continuing our running example:

  Only one swap operation happens, right before the fourth gate, making the total cost 1.
  So, $\map_3 = \map_2 = \map_1$.
  Before the fourth gate, the physical qubits $\physq_2$ and $\physq_3$
  are swapped, resulting in the map $\map_4$ that is the same as $\map_1$
  except that $\map_4(\logic_2) = \physq_3$ and $\map_4(\logic_3) = \physq_2$.
  \end{example}

\section{ Optimal QMR via maxsat}
\label{sec:alg}
In this section, we will present our approach for discovering an optimal solution of the \qmr problem via a reduction to \emph{maximum satisfiability} (\maxsat).
We begin by providing some background on \maxsat.

\subsection{MaxSAT Background}
\maxsat is the optimization analogue of the classical Boolean satisfiability problem, \sat. Before turning to our encoding, we will  define both of these problems.

\paragraph{The \sat problem}
In the \emph{satisfiability} problem (\sat), we are given a Boolean formula
and our goal is to find an assignment to the variables that makes the formula true---a \emph{model} of the formula.
We will use standard notation to denote Boolean operations: $\land$ (AND), $\lor$ (OR), $\neg$ (NOT), and $\rightarrow$ (implication).

\begin{example}
  Consider the following formula, where $a,b,c$ are Boolean variables:
  \[
  (\neg a \land b) \rightarrow c  
  \]
  This formula is \emph{satisfiable} because there is a model that
  makes it true. One such model is:
  \[
  \model = [a \mapsto \false,\ b \mapsto \true,\ c \mapsto \true]  
  \]
\end{example}
Given a formula $\form$, we will use $\model \models \form$ to denote
that $\model$ is a model of $\form$.
If $\form$ has no models, then it is \emph{unsatisfiable}.

\paragraph{The \maxsat problem} 
In the \maxsat problem, we are given two sets of Boolean formulas:
$\hard$ constraints and $\soft$ constraints.
Our goal is to find a single assignment $\model$ that is a model of
all hard constraints and as many soft constraints as possible.

\begin{example}
Consider the \maxsat problem with one hard constraint and two soft ones:
\begin{align*}
\hard & = \{\neg a \lor b \}\\ 
\soft &= \{ b,\ a\land \lnot b \}
\end{align*}
Since  $a\land \lnot b$ in $\soft$ is the negation of $\lnot a \lor b$, there is no $model$ of $\hard$ that is also a model 
of $\{a \land \lnot b \}$. Therefore the maximum number of formulas from $\soft$ that can evaluate to $\true$ is one. A solution is $\model = [a \mapsto \false, b \mapsto \true]$.
\end{example}

\subsection{MaxSAT Encoding of Optimal QMR}
We will now present our \maxsat encoding for optimally solving the \qmr problem.
Throughout, we fix a circuit $\circuit$ over logical qubits $\logics$ 
and a connectivity graph $\graph = (\phys,\edges)$.

Our encoding will define a set of hard constraints and a set of soft constraints, $(\hard,\soft)$, constituting a \maxsat instance. A solution to this \maxsat instance yields an optimal solution to the \qmr problem, specifically, (1) an optimal map sequence, $\seq{\map_1,\ldots,\map_{|\circuit|}}$, 
and (2) a sequence of \SWAPs before every two-qubit gate to perform routing.
The soft constraints aim to minimize the number of inserted \SWAPs.

Our encoding uses two sets of Boolean variables:
the $\vmap$ variables, which represent the sequence of maps $\seq{\map_1,\ldots,\map_{|\circuit|}}$,
and the $\vswap$ variables, which represent where \SWAPs are inserted
in the circuit.
In what follows, we describe our constraints in a semi-formal manner with examples and refer to \cref{fig:const} for the complete formalization.

\begin{figure}[t!]
  \small
  \begin{mdframed}[backgroundcolor=gray!10,linecolor=blue!0]
    \subsubsection*{Mapping constraints}
    \hfill\\

    \paragraph{\emph{Hard A}: Maps are injective functions}
    For every gate $g_k$ in the circuit, every logical qubit $\logic$, and every pair of distinct physical qubits $\physq,\physq'$,
    we add the following hard constraint:
    \begin{equation*}
      \vmap(\logic,\physq,k) \rightarrow \lnot \vmap(\logic,\physq',k)
    \end{equation*}
    Similarly,  for every $g_k$, every pair of distinct logical qubits, $\logic,\logic'$, and every physical qubit $\physq$, we add the following hard constraint:
    \begin{equation*}
      \vmap(\logic,\physq,k) \rightarrow \lnot \vmap(\logic',\physq,k)
    \end{equation*}

    \paragraph{\emph{Hard B}: Executing two-qubit gates}
  
For every two-qubit gate $g_k(\logic,\logic')$, we add the following hard constraint:
\begin{equation*}
\bigvee_{(\physq,\physq') \in \edges} (
\vmap(\logic,\physq,k) \land \vmap(\logic',\physq',k))
\end{equation*}
\hrule 
\vspace{1em}

\subsubsection*{Routing constraints}
\hfill\\

\paragraph{\emph{Hard C}: Only one swap}
For the $k$th gate and for its $i$th \SWAP, we add the hard constraint:
\begin{equation*}
 \bigvee_{(\physq,\physq') \in \edges'} \left(\vswap(\physq,\physq',k,i)
 \land \cEdge(\physq, \physq')\right)
\end{equation*}

where 
\[ \cEdge(p,p') \triangleq \bigwedge_{(r,r') \in \edges' \setminus\{(\physq,\physq')\}} \lnot\vswap(r,r',k, i)\] 

Here $\edges' = \edges \cup \{(\physq_0,\physq_0)\}$, a synthetic edge used
to denote a no-op \SWAP.

\paragraph{\emph{Hard D}: The effect of \SWAPs}
For every gate $g_k$ and sequence $S$ of swaps of physical qubits, $s(\physq_0,\physq_0'), \ldots, s(\physq_{n - 1},\physq_{n - 1}')$, we add the following hard constraint:  
\begin{equation*}
  \left(\bigwedge_{0 \leq i < n} \vswap(\physq_i,\physq_i',i,k)\right) \rightarrow \cSeff(S)
\end{equation*}
where 
\[ \cSeff(S) \triangleq  \bigwedge_{\substack{\logic \in \logics, \\phys \in \phys}} \left( \vmap(\logic,\physq,k-1) \leftrightarrow \vmap(\logic, \pi(S,\physq), k) \right)\]
\hrule 
\vspace{1em}

\subsubsection*{Soft constraints}
\hfill\\

\paragraph{\emph{Soft}: Minimize the number of \SWAPs}
For every gate $k$, we add the following soft constraint:
\begin{equation*}
   \vswap(\physq_0, \physq_0, k)
\end{equation*}
  \end{mdframed}
  \caption{Formalization of our \maxsat encoding}
  \label{fig:const}
  \end{figure}

\subsubsection{Mapping Constraints}
We start by describing the constraints that specify that
our map sequence is valid.

We will use the Boolean variable
$\vmap(\logic,\physq,k)$
to denote that, for a logical qubit $\logic \in \logics$
and a physical qubit $\physq \in \phys$, $\logic$ maps to $\physq$ 
right before the $k$th gate of the circuit.
In other words, if $\vmap(\logic,\physq,k)$ is assigned $\true$, then this means that $\map_k(\logic) = \physq$.

  \begin{example}\label{ex:vmap}
    Recall our running example from \cref{fig:example}.
    The initial map, $\map_1$, shown in \cref{fig:example}(bottom),
    maps $\logic_0$ to $\physq_1$.
    This is represented in our encoding by assigning the variable $\vmap(\logic_0,\physq_1,1)$ to $\true$ in the solution to the \maxsat constraints.
    Similarly, all other variables $\vmap(\logic_0,\physq_i,1)$, where $i\neq 1$,
    are set to $\false$, because $\logic_0$ can only be mapped to a single physical qubit.
  \end{example}

  \paragraph{\emph{Hard A}: Maps are injective functions}
Our first set of hard constraints (formalized in \cref{fig:const})
specify that our $\vmap$ variables model injective functions.
Following \cref{ex:vmap}, such constraints ensure that we cannot 
set both $\vmap(\logic_0,\physq_1,1)$ and $\vmap(\logic_0,\physq_2,1)$
to $\true$ in a solution of the \maxsat constraints.
Additionally, we cannot map different logical qubits to the same physical qubit.

\paragraph{\emph{Hard B}: Executing two-qubit gates}
Our second set of hard constraints specify that for each two-qubit gate in the  circuit, the two logical qubits it acts on are mapped to adjacent physical qubits.

\begin{example}
In our running example, the first gate is a \cnot over $q_0$ and $q_1$.
Therefore, we should initially map $q_0$ and $q_1$ to adjacent physical qubits.
One way to satisfy this is to set $\vmap(\logic_0,\physq_0,1)$ and $\vmap(\logic_1,\physq_1,1)$ to $\true$, since $(\physq_0,\physq_1)$ is an edge in the connectivity graph.
However, we cannot satisfy this constraint by setting $\vmap(\logic_0, \physq_0,1)$ and $\vmap(\logic_1,\physq_3,1)$ to $\true$, since there is no edge connecting physical qubits $(\physq_0,\physq_3)$.
\end{example}

\subsubsection{Routing Constraints}
We now describe the routing constraints.
The key idea is that right before a two-qubit gate, $g(\logic,\logic')$,
we want to insert a sequence of \SWAPs to ensure that the two logical qubits, $\logic$ and $\logic'$, are mapped to 
adjacent physical qubits.

Suppose, for illustration, that the $k$th gate in the circuit
is a \cnot over $q_0$ and $q_1$.
Right before this \cnot, we allow our encoding to insert  \emph{up to} $n$
\SWAPs. We can think of this through the lens of \emph{program sketching}~\cite{solar2008program},
where we don't know which qubits to swap before the $k$th gate,
so we add up to $n$ \SWAPs with unknown parameters (denoted with $\bullet$ below)
and the goal of our encoding is to \emph{discover} those parameters.
\begin{align*}
  \texttt{swap}& \ \bullet,\bullet\\
  \cdots &\\
  \texttt{swap}& \ \bullet,\bullet\\
  \texttt{cx}& \ q_0,q_1 
\end{align*}

Specifically, for every inserted \SWAP with unknown parameters,
we create a number of Boolean variables denoting every possible instantiation
of the parameters.
Formally, the Boolean variable $\vswap(\physq,\physq',k,i)$ denotes that the $i$th $\SWAP$ inserted before gate $k$ is over physical qubits $\physq$ and $\physq'$. 
(If both parameters are set to the same qubit, then the \SWAP is considered a no-op.)

\begin{example}
Suppose $n=1$---i.e., we only allow up to 1 \SWAP before a two-qubit gate---and we have a device with only two physical qubits, $\physq_0$ and $\physq_1$, with an edge between them.
For the fourth gate in the circuit,
we will have the following set of Boolean variables:
$$\{\vswap(\physq_0,\physq_0,4,1), \vswap(\physq_0,\physq_1,4,1)\}$$
If the first variable is set to $\true$ in a solution  to the \maxsat constraints, then no \SWAP is inserted before the fourth gate (no-op);
if the second variable is set to $\true$, then  
a \SWAP operation is inserted that swaps $\physq_0$ and $\physq_1$.
\end{example}

\paragraph{\emph{Hard C}: Only one swap}
As indicated by the above example, for a specific \SWAP with unknown parameters,
only one of its associated Boolean variables can be set to $\true$,
since there's only one possible instantiation of its parameters.
This is enforced by a standard \emph{only-one} hard constraint~\cite{gent2004new}.

\paragraph{\emph{Hard D}: The effect of \SWAPs}
The most involved routing constraint is encoding the effect of a sequence of \SWAPs on the map sequence.
Specifically, we have to encode how the inserted \SWAPs transform 
map $\map_{k-1}$ into $\map_k$.

We define a function $\pi(S,\physq)$ that specifies the effect of a sequence of \SWAPs $S$ on a physical qubit $\physq$, i.e., which qubit $\physq$ gets routed to after executing the swaps in $S$.
The following example provides a simple illustration:

\begin{example}
For the sequence $S$ that just swaps $\physq$ and $\physq'$,
we have $\pi(S,\physq) = \physq'$.
For the sequence $S$ that swaps $\physq$ and $\physq'$ and then $\physq'$ and $\physq''$, we have $\pi(S,\physq) = \physq''$.    
\end{example}

The final set of hard constraints (Hard D) encodes the effect of every possible sequence $S$ of $n$ \SWAPs.
While the number of sequences is exponential in $n$, in practice, we have experimentally found that a small constant suffices for finding optimal solutions (\cref{sec:eval}).

\subsubsection{Soft Constraints and Optimality}
We have described all the required hard constraints.
Finally, we define a set of soft constraints with the goal of minimizing the number of inserted $\SWAPs$.
Informally, we want to ensure that as many \SWAPs are no-ops as possible.
So, we maximize the number of $\true$ Boolean variables of the form $\vswap(\physq,\physq,k)$.

\cref{fig:const} fully formalizes all of the hard and soft constraints that our encoding generates.
So, if we are given a circuit $\circuit$ and a connectivity graph $\graph$,
we can use our encoding to generate a \maxsat instance $(\hard,\soft)$
whose solution results in an optimal \qmr solution.

Given a model $\model \models (\hard,\soft)$, we can extract a valid map sequence from the assignments of the Boolean variables of the form $\vmap(\logic, \physq, k)$ by setting $\map_k(q) = p$
exactly when $\model$ assigns $\vmap(\logic, \physq, k)$ to $\true$.

This following theorem states optimality of our  solutions:
\begin{Theorem} 
  \label{OptModelsareOptMaps}
    Let  $\model$ be a solution for $(\hard,\soft)$.
    Let $\seq{\map_1,\ldots,\map_{|C|}}$ be the map sequence corresponding to $\model$ as described above.
    Then, $\seq{\map_1,\ldots,\map_{|C|}}$ is an optimal solution of the \qmr problem (as per \cref{eq:opt}).
 \end{Theorem}

In the theorem above, we make the assumption that $n$ (the number of \SWAPs allowed before each \cnot) is set to the \emph{diameter} of the connectivity graph.
This ensures that we can always bring any two qubits into adjacent positions.

\paragraph{\diff{Encoding size}}
\diff{If $n$, the number of \SWAPs allowed before each \cnot, is held constant (see \cref{sec:eval}), the \maxsat encoding from \cref{sec:alg} scales polynomially with the size of the input circuit and architecture. In particular, a naive implementation requires $O(|C| \cdot |\edges|)$ variables and $O(|\phys|^2 \cdot |\logics| \cdot |C|)$ constraints (Hard A is the dominating term). However, using a standard ``only-one'' encoding~\cite{gent2004new}, we need only 
$O(|\phys| \cdot |\logics| \cdot |C|)$ constraints.
This is a more compact representation than \mqtex \cite{wille2019mapping}
and matches the asymptotic behavior of a subsequent \smt approach, \olsq \cite{tan2020optimal}.
While the number of constraints is roughly the same as \olsq, our sketch-based view allows us to eschew the use of integer arithmetic, eliminating the expensive theory-lemma generation of an \smt solver.
}


\section{A Locally Optimal Relaxation}\label{sec:local}
Solving the \maxsat encoding presented in the previous section
results in an optimal \qmr solution.
However, this can be expensive in practice due to the complexity
of \maxsat.
In this section, we will present a relaxation that produces \emph{locally} optimal solutions.
Specifically, our approach \emph{slices} the circuit into a number of subcircuits and solves a \maxsat problem for each of them in sequence.
The result is that we need to solve a number of smaller \maxsat problems.

\paragraph{Slicing the circuit}
We can think of a circuit $\circuit$ as a sequence of subcircuits, or \emph{slices}, $\seq{\circuit_0, \cdots, \circuit_{s}}$.
Recall \cref{fig:example2}, which shows a circuit viewed as two slices.
We will now demonstrate how to solve \qmr by solving \maxsat constraints
for each slice in isolation.

We do this iteratively, starting with $\circuit_0$ and going through the rest of the slices.
First, for $\circuit_0$, we simply generate a \maxsat instance $(\hard_0,\soft_0)$ and solve it
as described in the previous section.
This results in a model $\model_0$.
Then for every slice $\circuit_i$, where $i > 0$, we run the following procedure:

\begin{enumerate}
  \item Generate \maxsat constraints $(\hard_i,\soft_i)$ for  $\circuit_i$
  \item 
  For every variable $\vmap(q,p,|C_{i-1}|)$ set to $\true$ in $\model_{i-1}$,
  add $\vmap(q,p,1)$ to $\hard_i$.
  \item Solve $(\hard_i,\soft_i)$, generating a model $\model_i$
\end{enumerate}

The interesting step here is step 2, which connects the final mapping from slice ${i-1}$ with the initial mapping from slice $i$.
Specifically, we add the final mapping from slice ${i-1}$ as hard constraints 
on the initial mapping for slice $i$.

\begin{figure}[t]
  \centering
  \includegraphics[scale=1.4]{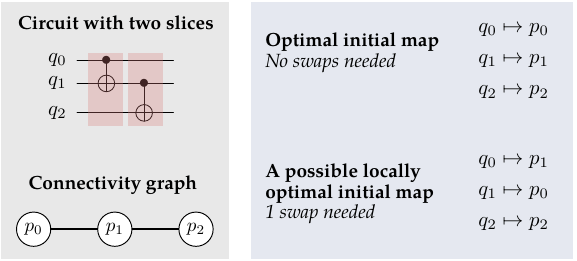}
  \caption{Example demonstrating local relaxation}
  \label{fig:localex}

\end{figure}

\begin{example}
Consider the circuit and connectivity graph in \cref{fig:localex}.
If we solve \qmr using the \maxsat encoding, a possible optimal initial map is the one that maps $\logic_i$ to $\physq_i$, as shown on the right, which requires no \SWAPs to be inserted.

Suppose, however, that we slice the circuit into two slices, as highlighted.
Solving the first slice might, for example, result in the map shown on the bottom right, with no swaps. This is optimal for the slice, but sub-optimal overall, since now we need to insert a \SWAP between the two gates.
Specifically, we will need to swap $\physq_0$ and $\physq_1$ (or $\physq_2$).
\end{example}

\paragraph{Backtracking}
If our constraints allow a number of swaps $n$ less than the diameter of the connectivity graph,
then we could generate unsatisfiable formulas for some slices.
In such cases, we backtrack to the previous slice and ask the \maxsat solver
to generate a different final mapping.
Backtracking from a mapping involves adding the negation of its corresponding satisfying assignment (previously returned by the \maxsat solver) as an additional hard constraint to ensure that the \maxsat solver does not return the same mapping.
After this additional constraint is added, backtracking is performed by re-invoking the solver.

\begin{example}
Consider again the map on the bottom right in \cref{fig:localex}. Let us suppose that we want to
backtrack and find a different mapping from this one for the first slice.
In order to guarantee that we do not return
the same mapping again when re-invoking the \maxsat solver, we add the following hard constraint:
$$\neg \big( \vmap(\logic_0, \physq_1, 2) \land
              \vmap(\logic_1, \physq_0, 2) \land
              \vmap(\logic_2, \physq_2, 2) \big)$$
This constraint is exactly the negation of the encoding of the mapping that we wish to exclude.
\end{example}


\section{Exploiting cyclic circuits}\label{sec:cyclic}

For some quantum algorithms, the quantum circuits 
have a repeated structure, applying the same subcircuit multiple times.
We call such circuits \emph{cyclic} circuits.

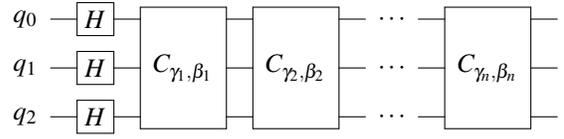
\begin{figure}[t]
  $$\Qcircuit @C=1em @R=0.6em 
  { \lstick{q_0}      & \gate{H} & \multigate{2}{\circuit_{\gamma_1,\beta_1}} &   \multigate{2}{\circuit_{\gamma_2,\beta_2}} &\qw &  \cdots && \multigate{2}{\circuit_{\gamma_n,\beta_n}} & \qw\\
  \lstick{q_1}      & \gate{H} & \ghost{\circuit_{\gamma_1,\beta_1}} &   \ghost{\circuit_{\gamma_2,\beta_2}} &\qw &  \cdots && \ghost{\circuit_{\gamma_3,\beta_3}} & \qw\\
  \lstick{q_2}      & \gate{H} & \ghost{\circuit_{\gamma_1,\beta_1}} &   \ghost{\circuit_{\gamma_1,\beta_1}} &\qw &  \cdots && \ghost{\circuit_{\gamma_1,\beta_1}} & \qw\\
  }$$
    \caption{The cyclic structure of a \qaoa circuit}
    \label{fig:qaoa}
  \end{figure}

The canonical algorithm that results in a cyclic circuit 
is the \emph{quantum approximate optimization algorithm} (\qaoa).
\qaoa is a general procedure for obtaining approximate solutions to \textsc{np}-hard combinatorial problems such as determining a maximum cut in a graph.
This is a promising near-term application since it solves problems of general practical interest and can be performed in the presence of noise without error-correction.
The general structure of a \textsc{qaoa} circuit is shown in \cref{fig:qaoa}.
Notice how the same subcircuit $\circuit_{\gamma,\beta}$ repeats.\footnote{Every cycle uses different parameters, $\gamma,\beta$, but the structure of the circuit remains the same, which is what matters for \qmr. Also, the initial set of one-qubit gates ($H$) is irrelevant for \qmr.}

\paragraph{A relaxation for cyclic circuits}
For cyclic circuits, instead of solving a \maxsat encoding for the entire circuit, we can relax the problem and only consider the repeating subcircuit.
After finding an optimal solution for the subcircuit, we can \emph{extend} the solution to the entire circuit.
This results in a smaller \maxsat problem that can generally be completed faster than the entire circuit, at the expense of a loss in optimality.

Suppose we have a circuit $\seq{\circuit,\ldots,\circuit}$, where the same subcircuit $\circuit$ is repeated a number of  times, and a connectivity graph $\graph$.
We follow the following simple recipe for solving the \qmr problem;
the key idea (step 2 below) is to ensure that the final map, $\map_{|\circuit|}$, is the same as the initial map, $\map_{1}$,
enabling us to \emph{stitch} together two or more copies of $\circuit$:
\begin{enumerate}
\item Let $(\hard,\soft)$ be the \maxsat constraints for $(\circuit,\graph)$.
\item 
For every pair of logical and physical qubits, $\logic$ and $\physq$, add the following constraint to $\hard$:
\[
 \vmap(\logic,\physq,1) \leftrightarrow \vmap(\logic,\physq,|\circuit|) 
\] 
\item Solve $(\hard,\soft)$, generating a model $\model$.
\end{enumerate}

The resulting circuit, with the initial map and swaps from $\model$,
can now be repeated an arbitrary number of times.
This is because the map of logical to physical qubits is the same 
at the beginning and at the end.

\begin{figure}[t!]
  \centering
  \includegraphics[scale=1.4]{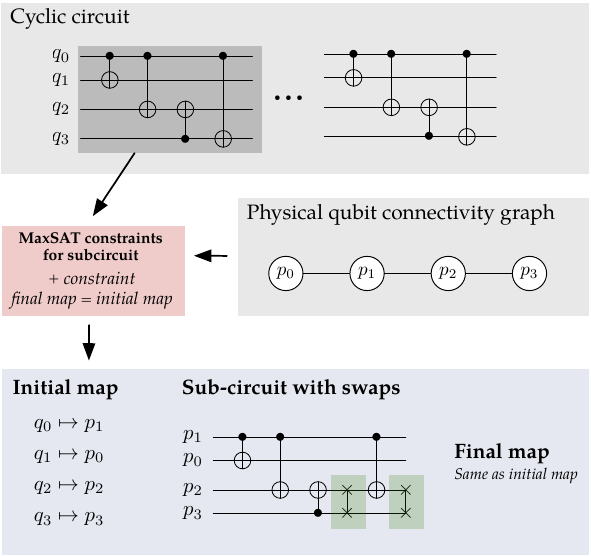}
  \caption{Illustration of our cyclic circuit approach}
  \label{fig:iterated}
\end{figure}

\begin{example}
  \cref{fig:iterated} revisits our running example,
  but with the same circuit iterated a number of times.
  Our cyclic-circuit approach in this case solves the same constraints as for \cref{fig:example}, except that we add the hard constraint that the initial and final maps are the same.
  This results in the final circuit with two \SWAPs, where the final \SWAP is inserted to reset the mapping to its initial state (i.e., to swap back $\physq_2$ and $\physq_3$).
  Now this subcircuit can be iterated any number of times on this physical connectivity graph.
\end{example}

This cyclic relaxation can be profitably combined with our local 
relaxation. Specifically, the \maxsat constraints generated in step 1 for subcircuit $C$
may be those that result from slicing it as described in \cref{sec:local}. Using such constraints, as we do in our evaluation (\cref{sec:eval}), allows our approach to handle larger subcircuits $C$.

\section{Implementation and Evaluation}
\label{sec:eval}

\paragraph{Implementation}
We implemented our approach in a tool we call \ours.\footnote{\url{https://github.com/qqq-wisc/satmap}} The value of $n$, the number of swaps allowed before each two-qubit gate, is set to 1. We experimentally determined $n=1$ is sufficient for near-optimal solutions. \ours generates \maxsat constraints and calls the \maxsat solver Open-WBO-Inc-MCS~\cite{joshi2018open} with default parameters.  This
solver provides the best known solution if it is interrupted before an optimal solution is found. 

We convert all \maxsat solutions to circuits. To ensure correctness of our \qmr solutions, we implemented an independent verifier. The verifier traverses a circuit, evaluating its effects on an initial map and checking that all two-qubit gates act on connected qubits.

Throughout this section, whenever we say \ours, we imply that the locally optimal relaxation (\cref{sec:local}) is performed. The cyclic relaxation is turned off by default and evaluated later on cyclic circuits.
The \emph{slice size} refers to the number of two-qubit gates to include in each slice in the local relaxation. We always run \ours at four slice sizes, 10, 25, 50, and 100, and  report the solution with the best cost.
Solution costs are in terms of \cnot gates added (\SWAP decomposes to 3 \cnots).

\paragraph{Benchmarks}
For evaluation, we used the set of benchmarks collected in \cite{zulehner2018efficient}.\footnote{\url{https://github.com/cda-tum/qmap/tree/main/examples}}
These 160 circuits were derived from the RevLib suite \cite{wille2008revlib} and programs written in the Quipper \cite{green2013quipper} and ScaffoldCC \cite{JavadiAbhari2014} quantum programming languages. 
They cover a wide spectrum of circuit size, ranging in number of qubits from 3 to 16, and in two-qubit gates from 5 to over 200,000. The median number of two-qubit gates among these benchmarks is 123. 

\begin{figure}[t]
  \centering
  \begin{subfigure}[b]{0.3\linewidth}
      \centering
      \begin{tikzpicture}[main/.style = {draw, circle, scale=0.58}] 
        \node[main] (1) {};
        \node[main] (2) [right of=1] {};
        \node[main] (3) [right of=2] {};
        \node[main] (4) [right of=3] {};
        \node[main] (5) [right of=4] {};
        \node[main] (6) [below of=1] {};
        \node[main] (7) [below of=2] {};
        \node[main] (8) [below of=3] {};
        \node[main] (9) [below of=4] {};
        \node[main] (10) [below of=5] {};
        \node[main] (11) [below of=6] {};
        \node[main] (12) [below of=7] {};
        \node[main] (13) [below of=8] {};
        \node[main] (14) [below of=9] {};
        \node[main] (15) [below of=10] {};
        \node[main] (16) [below of=11] {};
        \node[main] (17) [below of=12] {};
        \node[main] (18) [below of=13] {};
        \node[main] (19) [below of=14] {};
        \node[main] (20) [below of=15] {};
      
        \draw (1) -- (2);
        \draw (1) -- (6);
        \draw (2) -- (3);
        \draw (2) -- (7);
        \draw (3) -- (4);
        \draw (3) -- (8);
        \draw (4) -- (5);
        \draw (4) -- (9);
        \draw (5) -- (10);
        \draw (6) -- (7);
        \draw (6) -- (11);
        \draw (7) -- (12);
        \draw (7) -- (8);
        \draw (8) -- (13);
        \draw (8) -- (9);
        \draw (9) -- (14);
        \draw (9) -- (10);
        \draw (10) -- (15);
        \draw (11) -- (12);
        \draw (11) -- (16);
        \draw (12) -- (13);
        \draw (12) -- (17);
        \draw (13) -- (18);
        \draw (13) -- (14);
        \draw (14) -- (15);
        \draw (14) -- (19);
        \draw (15) -- (20);
        \draw (16) -- (17);
        \draw (17) -- (18);
        \draw (18) -- (19);
        \draw (19) -- (20);
      \end{tikzpicture}
      \caption{\archminus}
      \label{fig:tokyom}
  \end{subfigure}
  \hfill
  \begin{subfigure}[b]{0.3\linewidth}  
      \centering 
      \begin{tikzpicture}[main/.style = {draw, circle, scale=0.58}] 
        \node[main] (1) {};
        \node[main] (2) [right of=1] {};
        \node[main] (3) [right of=2] {};
        \node[main] (4) [right of=3] {};
        \node[main] (5) [right of=4] {};
        \node[main] (6) [below of=1] {};
        \node[main] (7) [below of=2] {};
        \node[main] (8) [below of=3] {};
        \node[main] (9) [below of=4] {};
        \node[main] (10) [below of=5] {};
        \node[main] (11) [below of=6] {};
        \node[main] (12) [below of=7] {};
        \node[main] (13) [below of=8] {};
        \node[main] (14) [below of=9] {};
        \node[main] (15) [below of=10] {};
        \node[main] (16) [below of=11] {};
        \node[main] (17) [below of=12] {};
        \node[main] (18) [below of=13] {};
        \node[main] (19) [below of=14] {};
        \node[main] (20) [below of=15] {};

        \draw (1) -- (2);
        \draw (1) -- (6);
        \draw (2) -- (3);
        \draw (2) -- (7);
        \draw (2) -- (8);
        \draw (3) -- (7);
        \draw (3) -- (4);
        \draw (3) -- (8);
        \draw (4) -- (5);
        \draw (4) -- (9);
        \draw (4) -- (10);
        \draw (5) -- (9);
        \draw (5) -- (10);
        \draw (6) -- (7);
        \draw (6) -- (11);
        \draw (6) -- (12);
        \draw (7) -- (11);
        \draw (7) -- (12);
        \draw (7) -- (8);
        \draw (8) -- (13);
        \draw (8) -- (9);
        \draw (8) -- (14);
        \draw (9) -- (13);
        \draw (9) -- (14);
        \draw (9) -- (10);
        \draw (10) -- (15);
        \draw (11) -- (12);
        \draw (11) -- (16);
        \draw (12) -- (13);
        \draw (12) -- (17);
        \draw (12) -- (18);
        \draw (13) -- (18);
        \draw (13) -- (17);
        \draw (13) -- (14);
        \draw (14) -- (15);
        \draw (14) -- (20);
        \draw (14) -- (19);
        \draw (15) -- (19);
        \draw (15) -- (20);
        \draw (16) -- (17);
        \draw (17) -- (18);
        \draw (18) -- (19);
        \draw (19) -- (20);
      \end{tikzpicture}
      \caption{\textsc{ibm} Q20 Tokyo}
      \label{fig:tokyo}
  \end{subfigure}
  \hfill
  \begin{subfigure}[b]{0.3\linewidth}   
      \centering 
      \begin{tikzpicture}[main/.style = {draw, circle, scale=0.58}] 
        \node[main] (1) {};
        \node[main] (2) [right of=1] {};
        \node[main] (3) [right of=2] {};
        \node[main] (4) [right of=3] {};
        \node[main] (5) [right of=4] {};
        \node[main] (6) [below of=1] {};
        \node[main] (7) [below of=2] {};
        \node[main] (8) [below of=3] {};
        \node[main] (9) [below of=4] {};
        \node[main] (10) [below of=5] {};
        \node[main] (11) [below of=6] {};
        \node[main] (12) [below of=7] {};
        \node[main] (13) [below of=8] {};
        \node[main] (14) [below of=9] {};
        \node[main] (15) [below of=10] {};
        \node[main] (16) [below of=11] {};
        \node[main] (17) [below of=12] {};
        \node[main] (18) [below of=13] {};
        \node[main] (19) [below of=14] {};
        \node[main] (20) [below of=15] {};
      
        \draw (1) -- (2);
        \draw (1) -- (6);
        \draw (1) -- (7);
        \draw (2) -- (3);
        \draw (2) -- (6);
        \draw (2) -- (7);
        \draw (2) -- (8);
        \draw (3) -- (7);
        \draw (3) -- (4);
        \draw (3) -- (8);
        \draw (3) -- (9);
        \draw (4) -- (5);
        \draw (4) -- (8);
        \draw (4) -- (9);
        \draw (4) -- (10);
        \draw (5) -- (9);
        \draw (5) -- (10);
        \draw (6) -- (7);
        \draw (6) -- (11);
        \draw (6) -- (12);
        \draw (7) -- (11);
        \draw (7) -- (12);
        \draw (7) -- (13);
        \draw (7) -- (8);
        \draw (8) -- (12);
        \draw (8) -- (13);
        \draw (8) -- (9);
        \draw (8) -- (14);
        \draw (9) -- (13);
        \draw (9) -- (14);
        \draw (9) -- (10);
        \draw (9) -- (15);
        \draw (10) -- (14);
        \draw (10) -- (15);
        \draw (11) -- (12);
        \draw (11) -- (16);
        \draw (11) -- (17);
        \draw (12) -- (16);
        \draw (12) -- (13);
        \draw (12) -- (17);
        \draw (12) -- (18);
        \draw (13) -- (18);
        \draw (13) -- (17);
        \draw (13) -- (19);
        \draw (13) -- (14);
        \draw (14) -- (15);
        \draw (14) -- (18);
        \draw (14) -- (20);
        \draw (14) -- (19);
        \draw (15) -- (19);
        \draw (15) -- (20);
        \draw (16) -- (17);
        \draw (17) -- (18);
        \draw (18) -- (19);
        \draw (19) -- (20);
      \end{tikzpicture}
      \caption{\archplus}
      \label{fig:tokyop}
  \end{subfigure}
  \caption{Variations of the \textsc{ibm} Q20 Tokyo graph} 
  \label{fig:archs}
\end{figure}
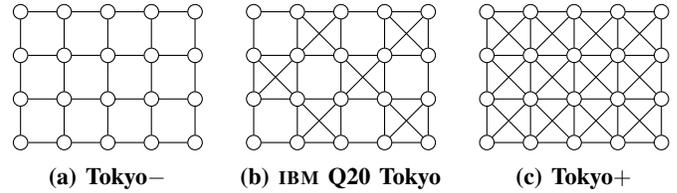

Except in the evaluation of Q4, the connectivity graph used is the \textsc{ibm} Q20 Tokyo architecture with 20 qubits, depicted in \cref{fig:tokyo}. This connectivity graph was chosen as the largest typically used for evaluation in related work \cite{li2019tackling, cowtan2019qubit, siraichi2019qubit}.
Benchmarks were evaluated on a cluster of Intel\textsuperscript{\textregistered} Xeon\textsuperscript{\textregistered} and AMD Opteron\texttrademark \xspace CPUs clocked an average of 2.5GHz.

\paragraph{Research questions}
We designed a set of experiments to answer the following research questions:
\begin{description}
\item[Q1] How does \ours compare to constraint-based techniques?
\item[Q2] How does \ours compare to heuristic approaches?
\item[Q3] What is the impact of local relaxation and cyclic circuit relaxation?
\item[Q4] How does architecture variation impact performance?
\item[Q5] \diff{What is the scalability vs. optimality tradeoff?}
\item[Q6] \diff{Can we use \ours with noise models?} 
\end{description}

\begin{figure}
  \centering
  \includegraphics[width=\linewidth]{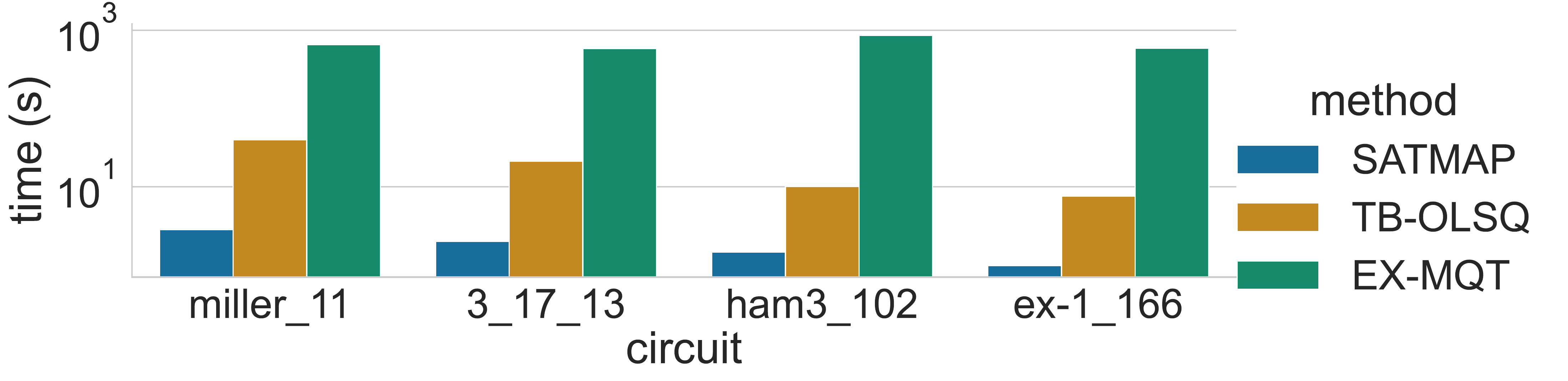}
  \caption{Log-scale runtime comparison of \mqtex, \olsq, and \ours on the set of benchmarks \mqtex solved}
  \label{fig:barchart_jku}
\end{figure}
\begin{figure*}[h]
  \includegraphics[width=\linewidth]{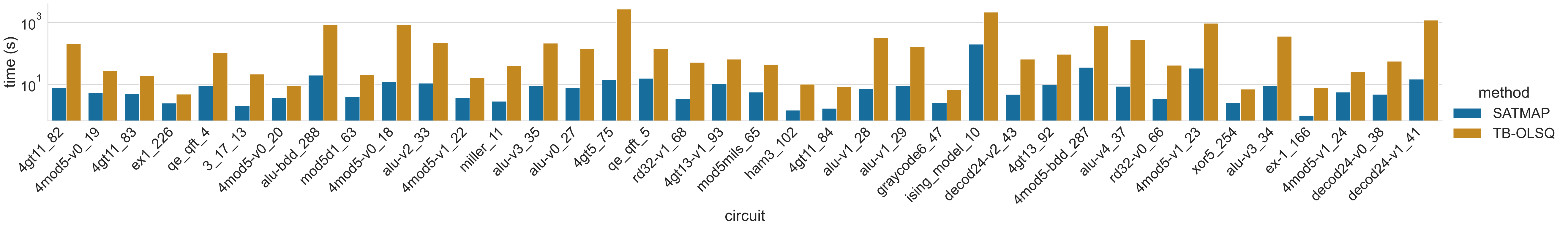}
  \caption{Log-scale runtime comparison of \olsq and \ours on the complete set of benchmarks \olsq solved}
  \label{fig:barchart_all}
\end{figure*}

\subsection*{Q1: Constraint-based approaches}


\paragraph{Experimental setup}
To address Q1, we compared \ours to the \smt-based tools \mqtex ~\cite{wille2019mapping} and \olsq ~\cite{tan2020optimal}. The latter tool takes the
relative execution time of each gate as input and can optimize several different 
objective functions. We set the execution time of each gate to 1 and the objective function to \SWAP minimization to match our definition of \qmr.
For each of the benchmarks, \ours was allotted 30 minutes of compilation time and the other tools were each allotted 1 hour of compilation time to be as fair 
as possible and account for any potential hidden overheads. Each tool was allotted 5\textsc{gb} of \textsc{ram} for each of the benchmarks.

\paragraph{Results}
None of the tools were able to provide a solution to the \qmr problem for all of the benchmarks within the time and memory restrictions. However, as \cref{cbased-num-solved} indicates, \ours handles a significantly higher proportion
of the complete set, solving 109/160 (68\%) of the benchmarks as compared to 4/160 (2.5\%) solved by \mqtex and 38/160 (24\%) solved by \olsq. The additional benchmarks solved by \ours include circuits with up to 598 two-qubit gates versus a maximum of 23 for \mqtex and 90 for \olsq.
\begin{table}
\centering
\small
  \begin{tabular}{c c c}
    \toprule
    Tool  & \# Solved (out of 160) & Largest circuit solved\\
    \midrule
    \mqtex  & 4 & 23 \\ 
    \olsq & 38 & 90 \\ 
    \ours & \textbf{109} & \textbf{598}\\
    \bottomrule
  \end{tabular}
  \caption{Comparison against constraint-based tools. Size of largest circuit solved is the number of two-qubit gates.}
  \label{cbased-num-solved}
\end{table}

Since these tools are all designed to provide optimal solutions,
there is no variation in quality of the solutions.\footnote{\olsq formulates \qmr in terms of time coordinates, which treats a broader class of circuits as equivalent, allowing solutions not considered by the other tools. Additionally, a minor relaxation in \olsq produces suboptimal solutions in rare cases. 
The difference in cost due to these considerations is less than one \SWAP in all cases.} Without sacrificing optimality, \ours significantly
outperformed the two other tools in terms of runtime. The mean improvement over \mqtex was about 400x and the mean improvement over \olsq was about 20x, as shown in  \cref{fig:barchart_jku} and \cref{fig:barchart_all}.

\textbf{Summary: \ours is significantly more scalable than \mqtex and \olsq. It finds solutions 400x and 20x faster (respectively) and can be applied to much larger circuits, up to 598 two-qubit gates.}

\begin{figure*}[h]
  \centering
  \begin{subfigure}[b]{0.3\textwidth}
      \centering
      \includegraphics[width=\textwidth]{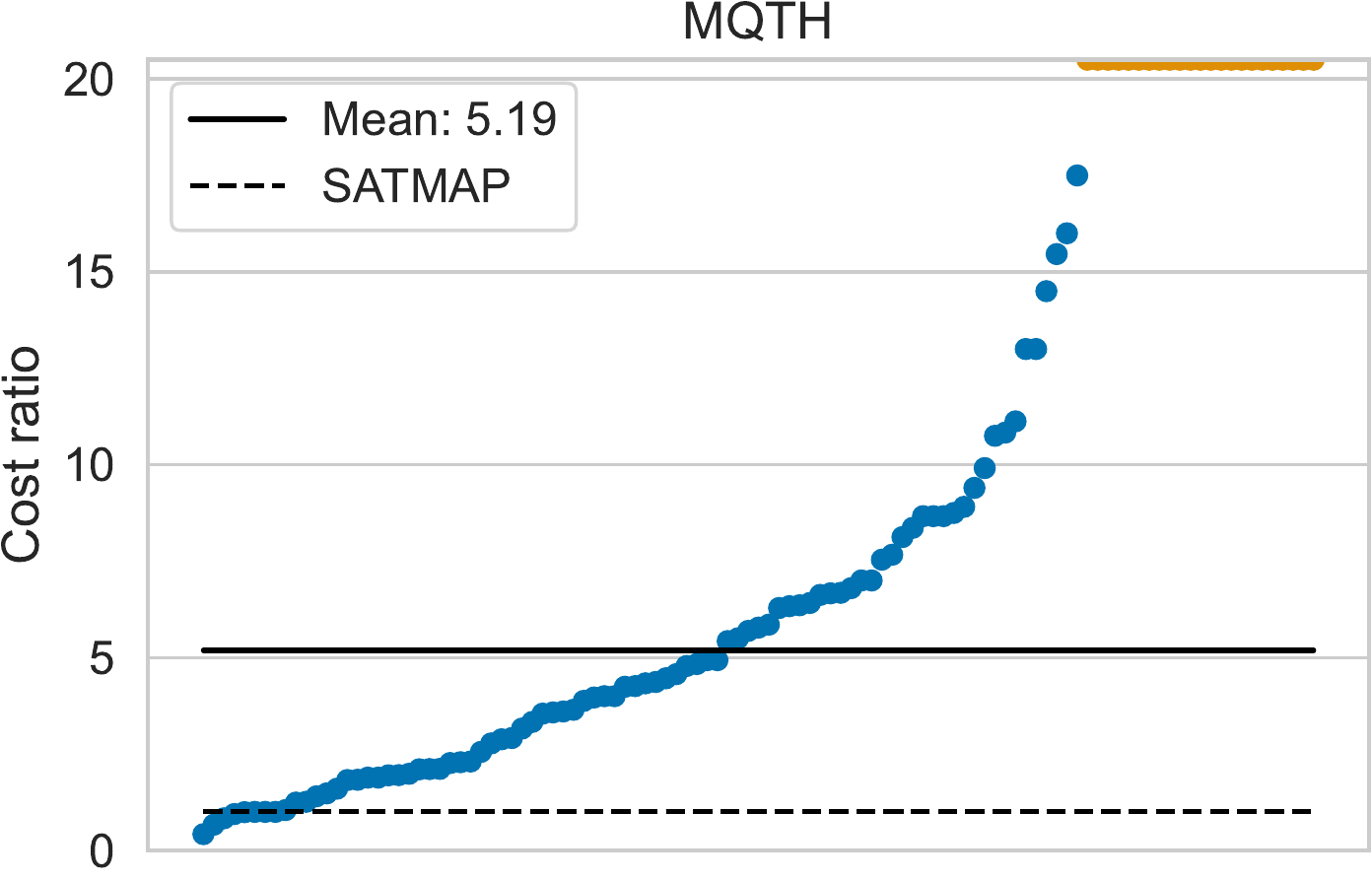}
  \end{subfigure}
  \hfill
  \begin{subfigure}[b]{0.3\textwidth}  
      \centering 
      \includegraphics[width=\textwidth]{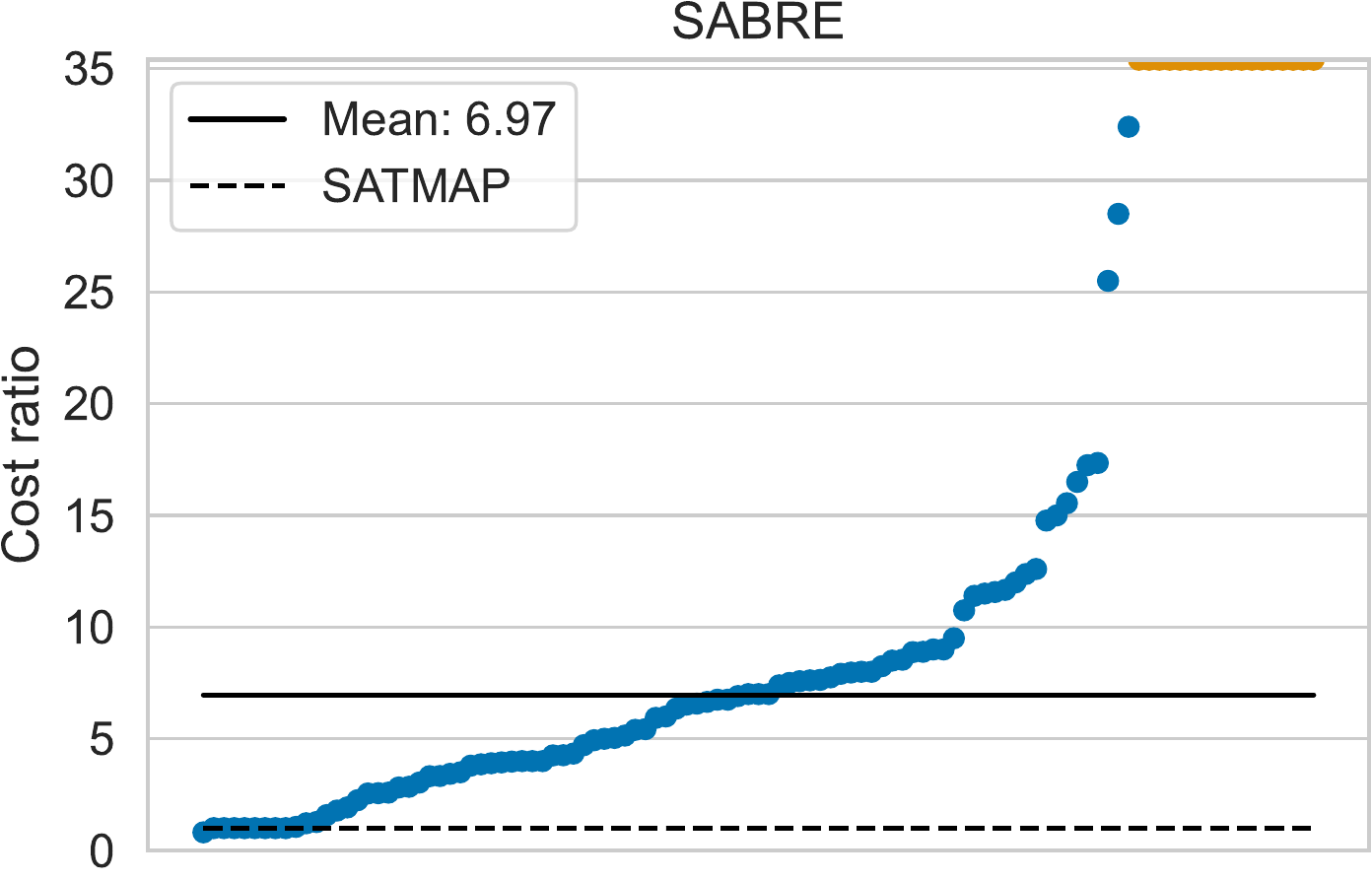}
  \end{subfigure}
  \hfill
  \begin{subfigure}[b]{0.3\textwidth}   
      \centering 
      \includegraphics[width=\textwidth]{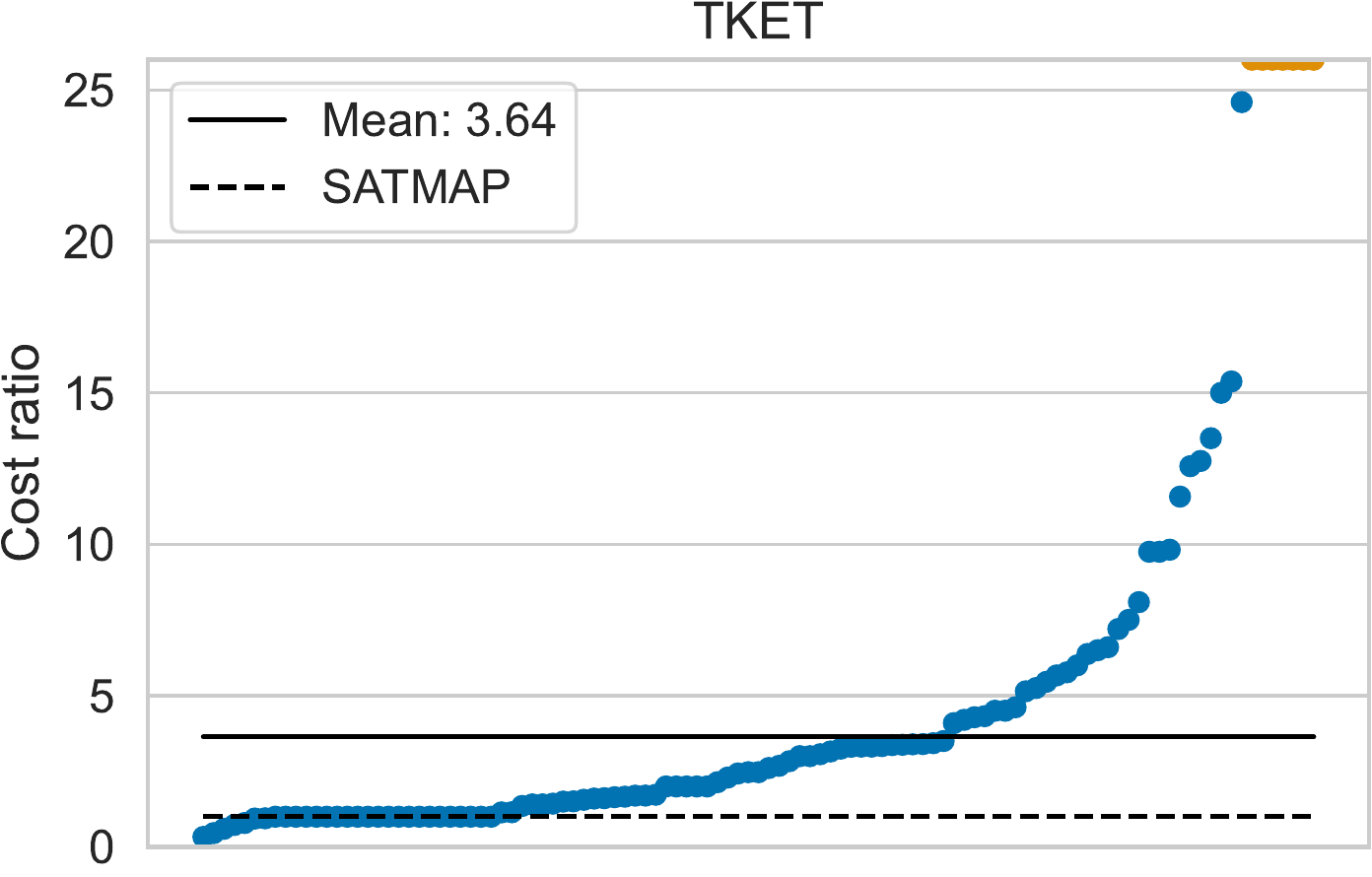}

  \end{subfigure}
  \caption
  { The cost of the solution produced by each heuristic tool divided by the cost of the solution produced by \ours. Points at the top of the plot represent benchmarks where \ours 
  added zero gates and the heuristic tool added a positive number, resulting in an undefined ratio. They are not included in the listed mean ratio.} 
  \label{fig:heuristic}
\end{figure*}

\subsection*{Q2: Heuristic approaches}
\paragraph{Experimental setup}
To address Q2, we compared \ours to the heuristic tools \mqth \cite{zulehner2018efficient}, \sabre \cite{li2019tackling}, and \tket \cite{cowtan2019qubit}. \mqth applies $\text{A}^*$ search to determine the optimal next mapping given the current one. \sabre and \tket both use heuristic ``scores'' to choose \SWAPs to apply to the qubits relevant to a particular topological layer or timestep of a circuit, with \sabre applying
a reversal procedure to determine a good initial mapping. These three were chosen to represent the state-of-the-art in terms of heuristic tools based on their widespread use in practice. Each is
relied upon as a part of industrial quantum compilation toolkits. Since \tket and \sabre involve some element of nondeterminism, we took the mean cost of 20 runs of the heuristic tools.
 As in the evaluation of Q1, \ours was allotted 30 minutes of compilation time and 5\textsc{gb} of \textsc{ram} for each of the benchmarks. The heuristic tools
are less resource intensive, so they solved all of the instances well within those runtime and memory bounds.

\paragraph{Results}
For the 109 benchmarks solved within the timeout by \ours, the resulting solutions were generally better than the heuristic tools. \cref{fig:heuristic} presents the \emph{cost ratio} on each benchmark:
the total number of gates added by the heuristic tools divided by the total number of gates added by \ours. For each heuristic tool, there are rare instances (fewer than 10 benchmarks) when the heuristic outperforms \ours due to application of the local relaxation or early
termination of the \maxsat solver, resulting in ratios less than 1. On average, \ours adds 5.2x, 7.0x, and 3.6x fewer gates than \mqth, \sabre, and \tket, respectively. For all heuristic tools, there was at least one instance where 
\ours produced a solution with over 15x fewer gates.
For about 14\% of benchmarks, \ours did not add any gates, compared to 0\%, 3\%, and 10\% for \mqth, \sabre, and \tket, respectively. Benchmarks where \ours added no gates and a heuristic tool added some gates are represented by the orange points at the top of the plot. Benchmarks where neither tool added gates have a cost ratio of 1.

\textbf{Summary: when \ours terminates, it gives much higher quality solutions than heuristic tools overall. It almost always reduces the total cost---up to 6.97x on avg.}

\subsection*{Q3: Impact of Relaxations}

\paragraph{Local relaxation}
We conducted experiments to determine the effect of the locally optimal relaxation (\cref{sec:local}) on performance in terms of execution time and cost. 
We tested the slice sizes, $10$, $25$, $50$, and $100$, against \nolocal, which is \ours with local relaxation disabled.

Small slice sizes produce easier \maxsat problems that can each individually be solved faster. However, restricting the ``view'' of the solver can lead to increased overall
solve time due to repeated backtracking. We can observe this tradeoff by comparing the number of instances for which a solution was found within the 30 minute timeout (shown in \cref{table:lr}). 

\begin{figure}[t]
  \centering
  \begin{subfigure}[b]{0.49\linewidth}
    \centering
    \includegraphics[width=\textwidth]{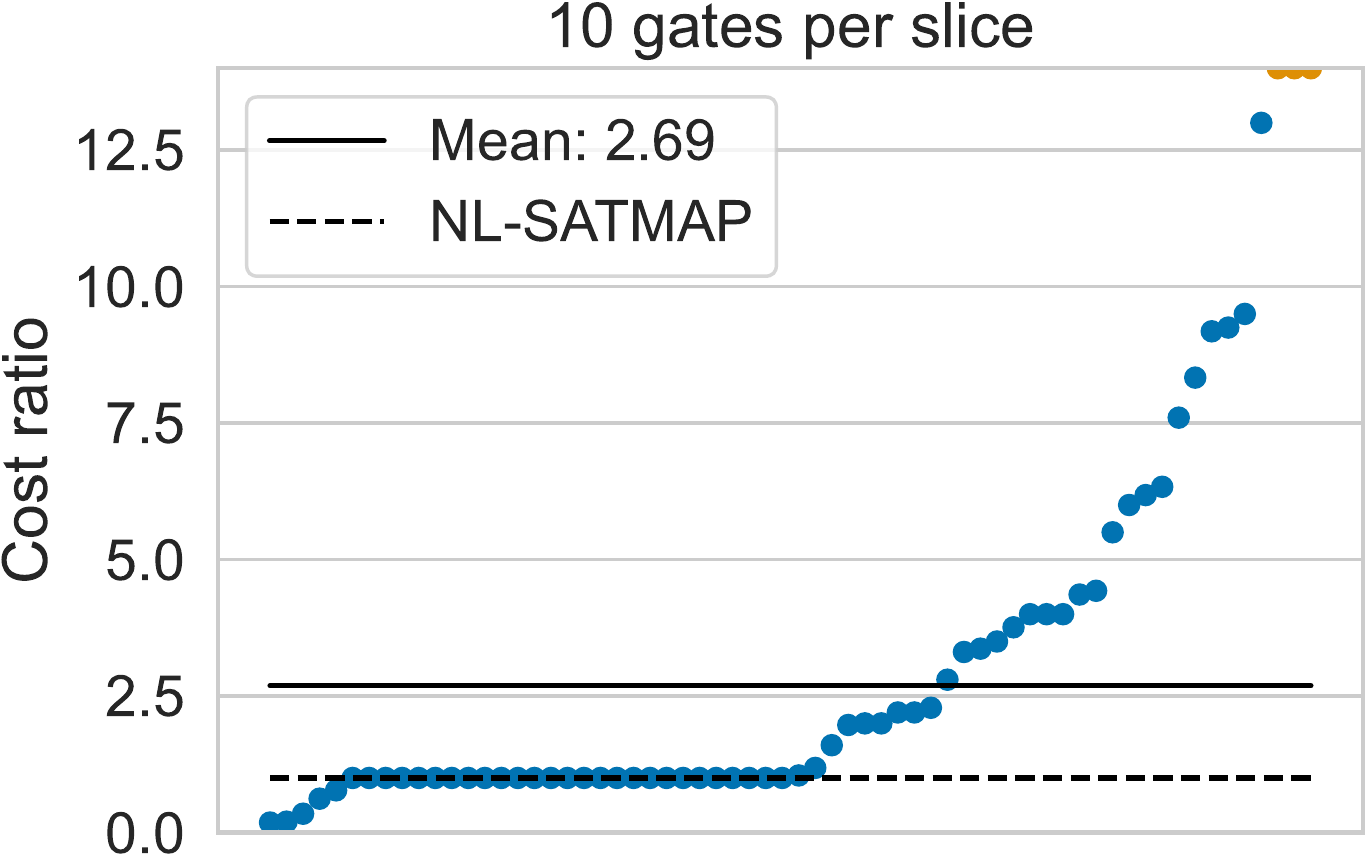}
  \end{subfigure}
  \hfill
  \begin{subfigure}[b]{0.49\linewidth}  
    \centering 
    \includegraphics[width=\textwidth]{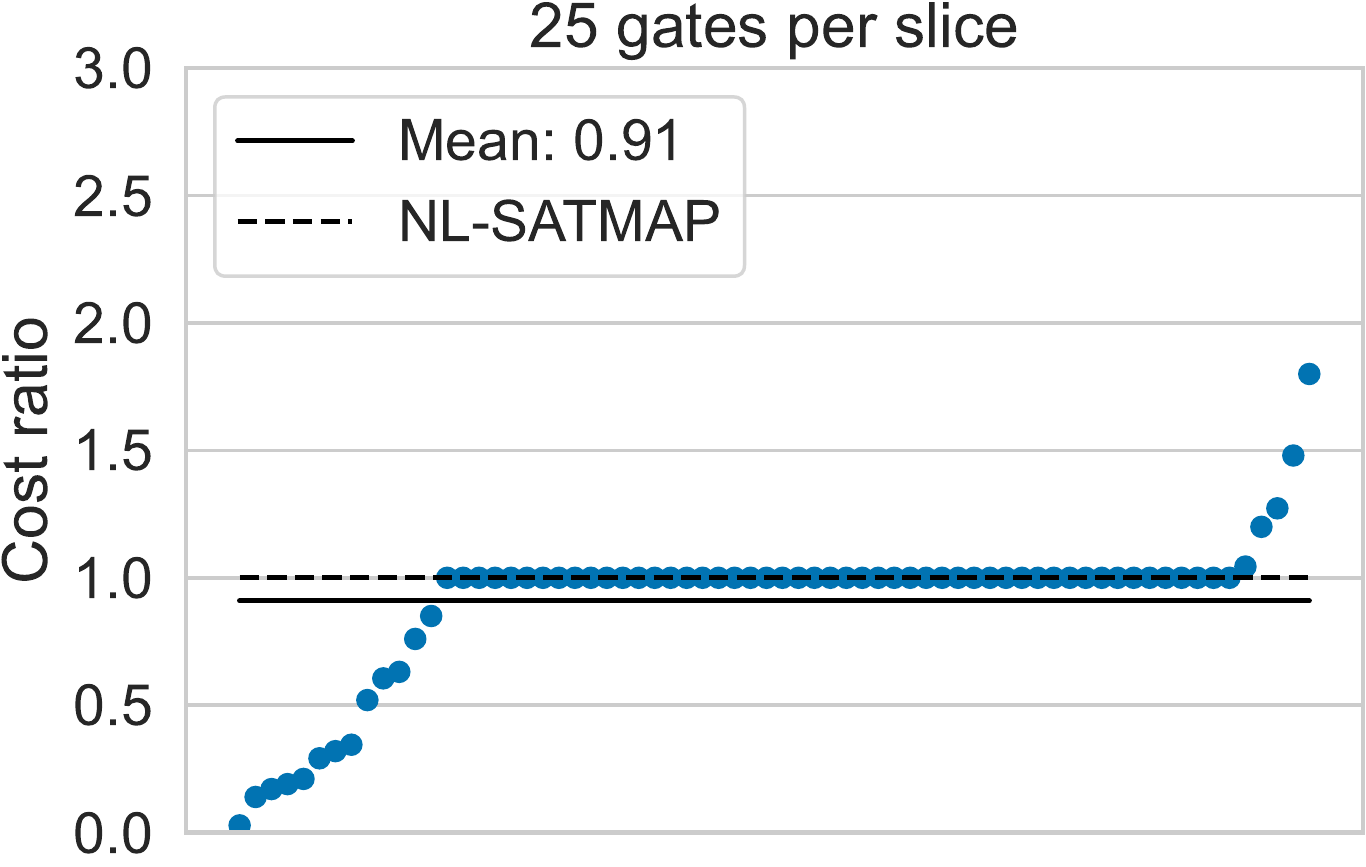}

  \end{subfigure}
  \vskip\baselineskip
  \begin{subfigure}[b]{0.49\linewidth}   
    \centering 
    \includegraphics[width=\textwidth]{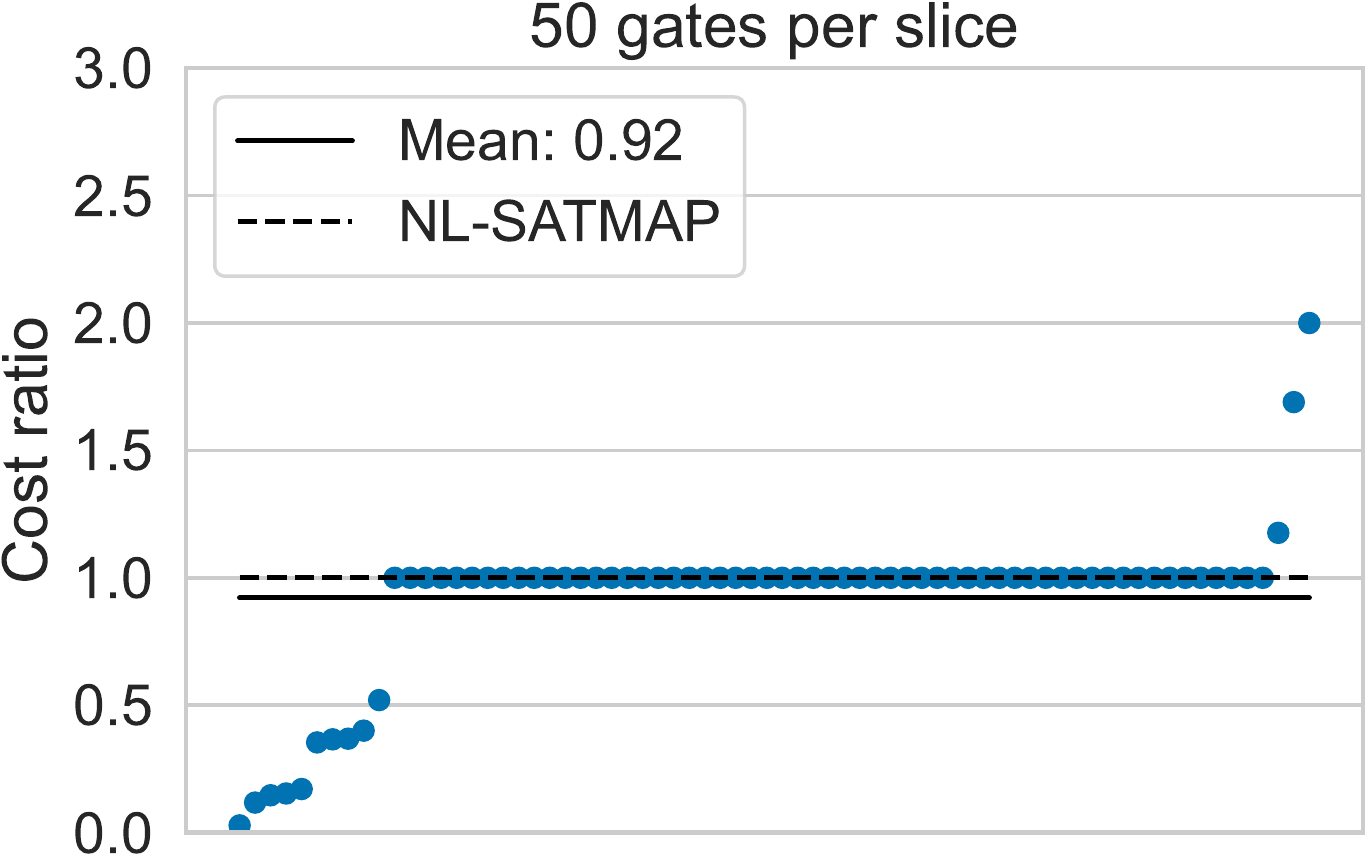}
  
  \end{subfigure}
  \hfill
  \begin{subfigure}[b]{0.49\linewidth}   
    \centering 
    \includegraphics[width=\textwidth]{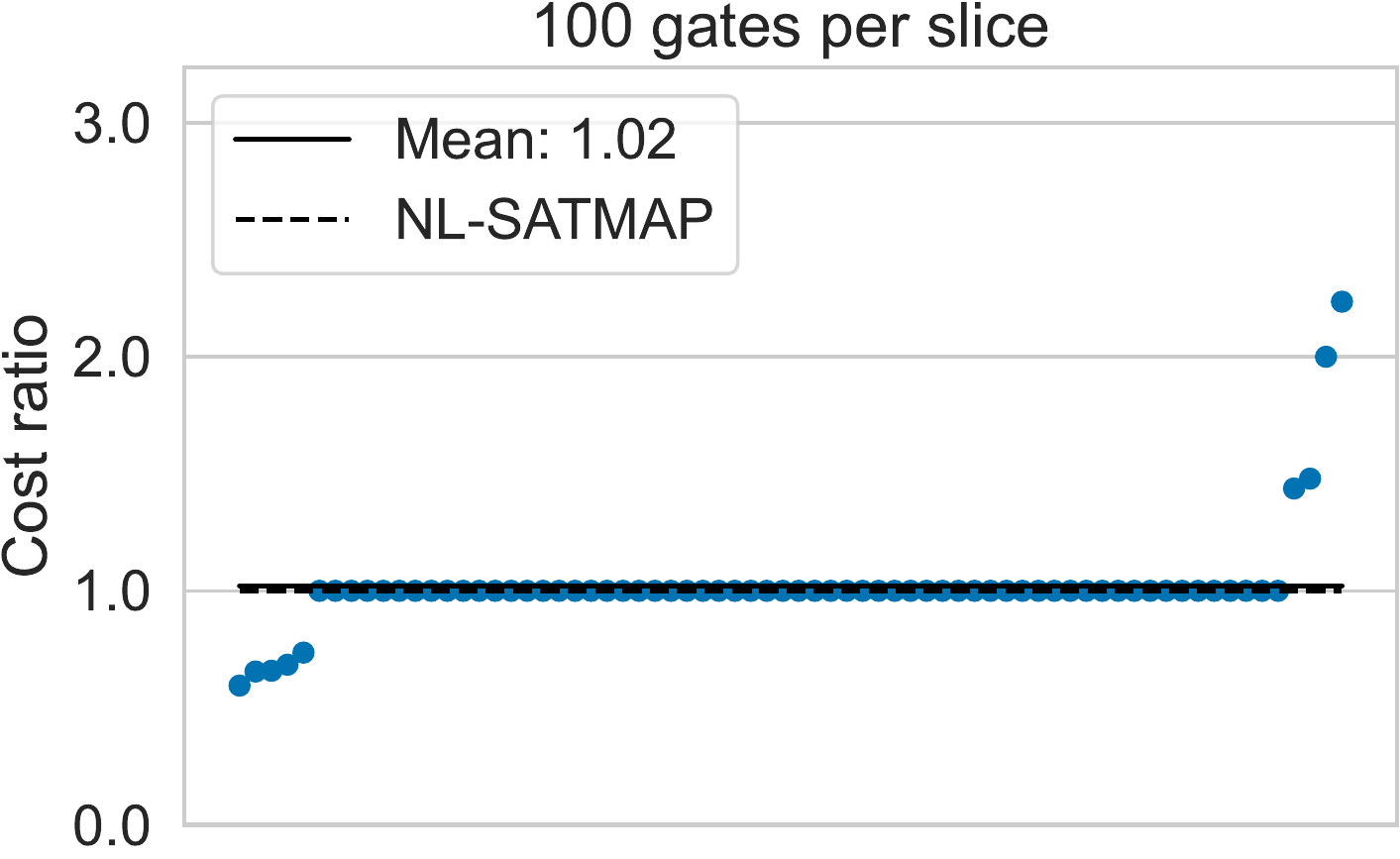}
  
  \end{subfigure}
  \caption{The cost of the solution produced by the different levels of local relaxation divided by the cost of the solution produced by \nolocal}
  \label{fig:rq3}
  \end{figure}

The situation is similar in terms of solution quality. For small slice sizes, local optima can diverge significantly from global optima. However, for moderate slice sizes, this effect is less pronounced. \cref{fig:rq3} presents the cost ratio of local relaxation levels: the total number of gates added by the tool with local relaxation divided by the total number of gates added by \nolocal. For a slice size of 10, \nolocal consistently produces better solutions, with an average cost ratio of 2.69. For larger slice sizes, the benchmarks where \nolocal discovers better solutions are
outnumbered by the benchmarks where the slow rate of convergence in a large solution space leads to worse solutions due to early termination. For example, with a slice size of 25, \nolocal produces a better solution in 5 out of 70 cases, but a worse solution in 13 out of 70. This results in a mean ratio of less than 1 (0.92).

\textbf{Summary: the local relaxation is a significant contributor to the performance of \ours. Appropriate application of local relaxation enables the use of constraint-based tools on large benchmarks, with little loss in solution quality for smaller ones. }

  \begin{table}[t]
    \centering
    \small
      \begin{tabular}{c c c}
        \toprule
        Slice size  & \# Solved (out of 160) & Largest circuit solved\\
        \midrule
         10 & 87 & 427 \\ 
        25 & \textbf{103} & \textbf{598} \\ 
        50 & 92 & \textbf{598}\\
        100 & 71 & 128\\
        \nolocal & 70 &  128 \\
        \bottomrule
      \end{tabular}
      \caption{Comparison of different levels of local relaxation in terms of instances solved}
      \label{table:lr}
  \end{table}

  \begin{table}[t]
    \centering
    \small
    \setlength{\tabcolsep}{2pt}
      \begin{tabular}{c c c c c}
        \toprule
        Tool  & \multicolumn{2}{c}{Main Set (total 160)} & \multicolumn{2}{c}{\qaoa (total 10)}\\
        & \# Solved & Largest solved & \# Solved & Largest solved\\
        \midrule
        \olsq & 38 & 90 & 0 & --\\
        \nolocal & 70 &  128  & 5 & 36\\
        \ours & 109 & 598 & 7 & 72  \\
        \cyclic & -- & -- & 10 & 96\\
        \bottomrule
      \end{tabular}
      \caption{\diff{Comparison between \olsq, \nolocal, \ours, and \cyclic}}
      \label{table:olsq_vs_us}
  \end{table}

\paragraph{Cyclic relaxation}
To evaluate the cyclic relaxation, we programmatically generated a standard \qaoa circuit for solving the maximum cut problem on 3-regular graphs,
parameterized by the number of qubits and the number of cycles (repetitions of the subcircuit $\circuit_{\gamma,\beta}$). 
We use \cyclic to denote \ours with the cyclic relaxation enabled.

We tested \ours, \cyclic, and \tket on \qaoa circuits with 6, 8, 10, 12 and 16 qubits, each with two and four cycles. 
The results  are presented in \cref{table:cr}. 
Missing entries correspond to timeouts. 

\cyclic solves all of the instances within the timeout, while \ours
cannot come up with a solution for circuits with 10 qubits or 16 qubits. 
When it comes to cost, for 6 and 8 qubits, \cyclic outperforms \ours, whereas for 10 and 12 qubits, the opposite is true.\footnote{Note that \sat solvers do not provide a monotonicity guarantee on performance with respect to increasing circuit size---e.g., we can solve a 12-qubit circuit with \ours but not a 10-qubit circuit. This is due to the search strategy employed by the underlying solver.} 
Additionally, except with 12 qubits, \cyclic determines a solution much more quickly than \ours. 
Neither \ours nor \cyclic has a clear advantage in terms of solution quality when both produce some solution. 
In some cases, \cyclic enables us to find a better solution than the best heuristic tool, \tket.
For instance, with 16 qubits, \cyclic solutions are $>3x$ better than \tket.

\textbf{Summary: the cyclic relaxation improves performance of our approach on cyclic circuits in three respects: (1) it renders larger circuits tractable (such as 16-qubit \qaoa), (2) it produces a solution faster, and (3) it produces better solutions within a fixed timeout for some circuits.}

\diff{
\paragraph{Breakdown of effects}  
\cref{table:olsq_vs_us} summarizes the effects of our encoding and relaxations, starting with \olsq as the baseline.
Without relaxations, \nolocal can solve 70 benchmarks, while \olsq only 38. 
With local relaxation, \ours can solve 109 benchmarks, with a largest circuit of size 598, compared to \olsq's 90.
We also see the same behavior in \qaoa benchmarks, with the local relaxation (\ours) and cyclic relaxations (\cyclic) allowing us to solve progressively more benchmarks. \olsq is unable to solve any of our \qaoa benchmarks within the allotted time.
}

\begin{table*}[t]
  \centering
  \setlength{\tabcolsep}{14pt}
  \renewcommand{\arraystretch}{1.25}
    \begin{tabular}{c c c c c c c c}
      \toprule
      \# Qubits & Cycles & \multicolumn{2}{c}{\cyclic} & \multicolumn{2}{c}{\ours} & \multicolumn{2}{c}{\tket} \\
       &  & Cost & Time & Cost & Time & Cost & Time \\
      \midrule
      \multirow{2}{0.5em}{6} & 2 & \textbf{12} & 130 & \textbf{12} & 1800 & \textbf{12} & $<0.1$\\
      & 4 & \textbf{24} & 130 & 60 & 203 & 42 & $<0.1$ \\
      \midrule
      \multirow{2}{0.5em}{8} & 2 & \textbf{12} & 361 & 18 & 954 & 21 & $<0.1$ \\
      & 4 & \textbf{24} & 361 & 63 & 372 & 30 & $<0.1$ \\
      \midrule
      \multirow{2}{1em}{10} & 2 & 84 & 253 & 54 & 1155 & \textbf{33} & 0.14 \\
      & 4 & 168 & 253  & -- & -- & \textbf{102} & 0.24 \\
      \midrule
      \multirow{2}{1em}{12} & 2 & 84 & 1800  & \textbf{21} & 261 & 48 & 0.33 \\
      & 4 & 168 & 1800  & 105 & 1800 & \textbf{87}  & 0.32\\
      \midrule
      \multirow{2}{1em}{16} & 2 & \textbf{24} & 288 & --  & --  & 78 &  0.30\\
      & 4 & \textbf{48} & 288 & -- & -- & 147 & 0.41  \\
      \bottomrule
    \end{tabular}
    \caption{Quality of solutions and runtime (s) of \cyclic, \ours, and \tket on \qaoa circuits}
    \label{table:cr}
  \end{table*}

\begin{figure*}[t!]
  \centering
  \begin{subfigure}[b]{0.3\textwidth}
    \centering
    \includegraphics[width=\textwidth]{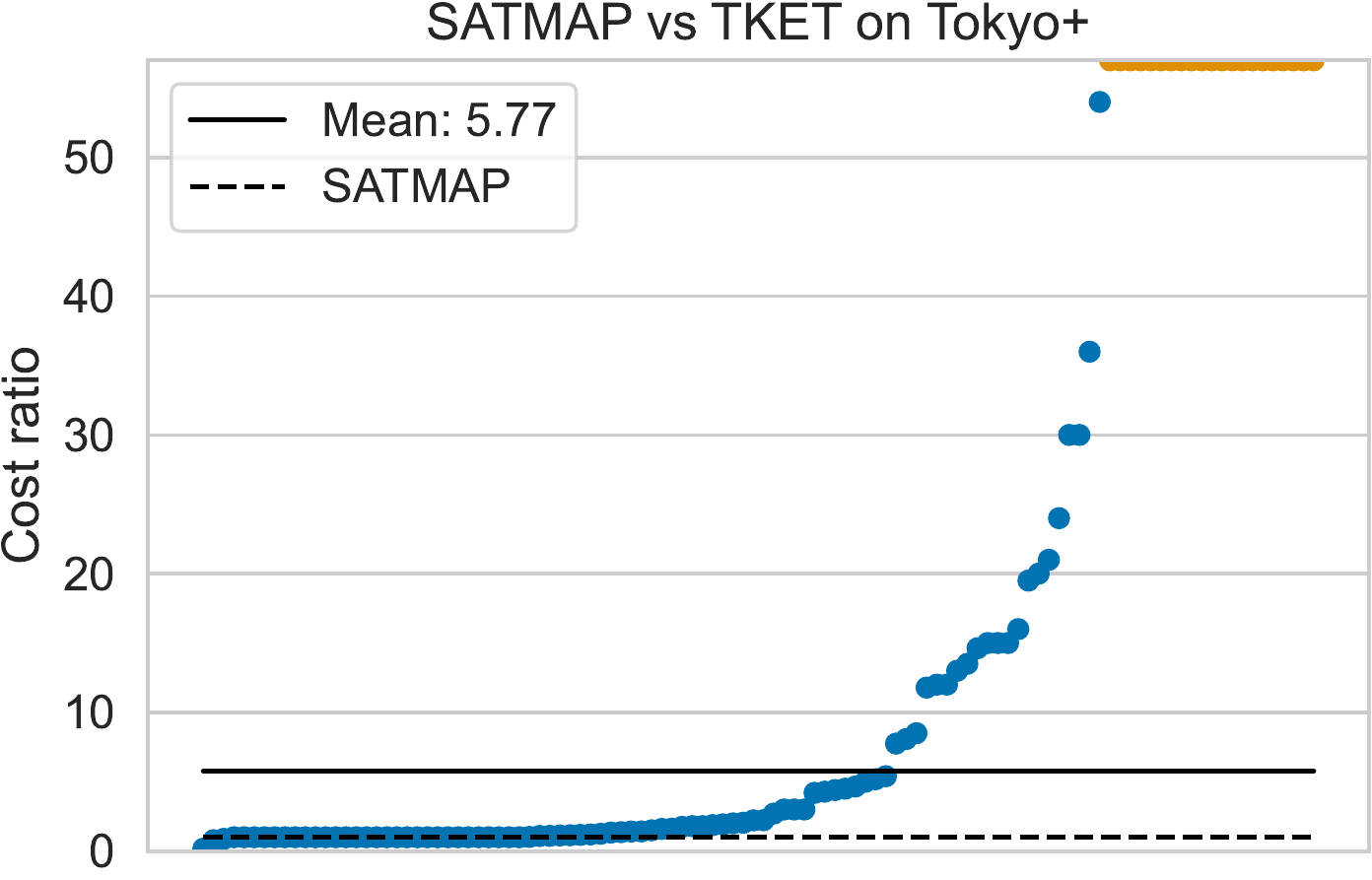}
  \end{subfigure}
  \hfill
  \begin{subfigure}[b]{0.3\textwidth}  
    \centering 
    \includegraphics[width=\textwidth]{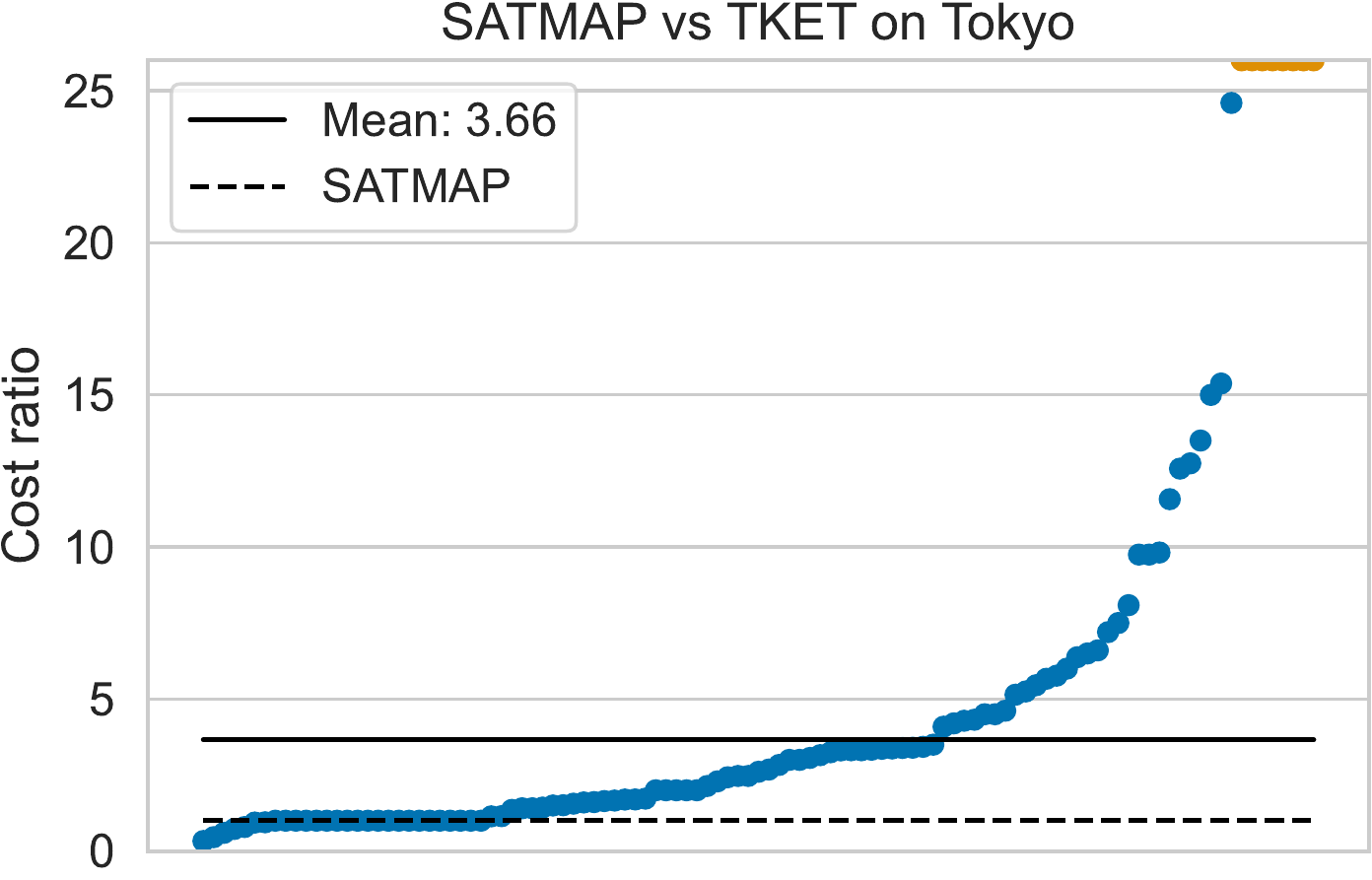}
  \end{subfigure}
    \hfill
  \begin{subfigure}[b]{0.3\textwidth}   
        \centering 
        \includegraphics[width=\textwidth]{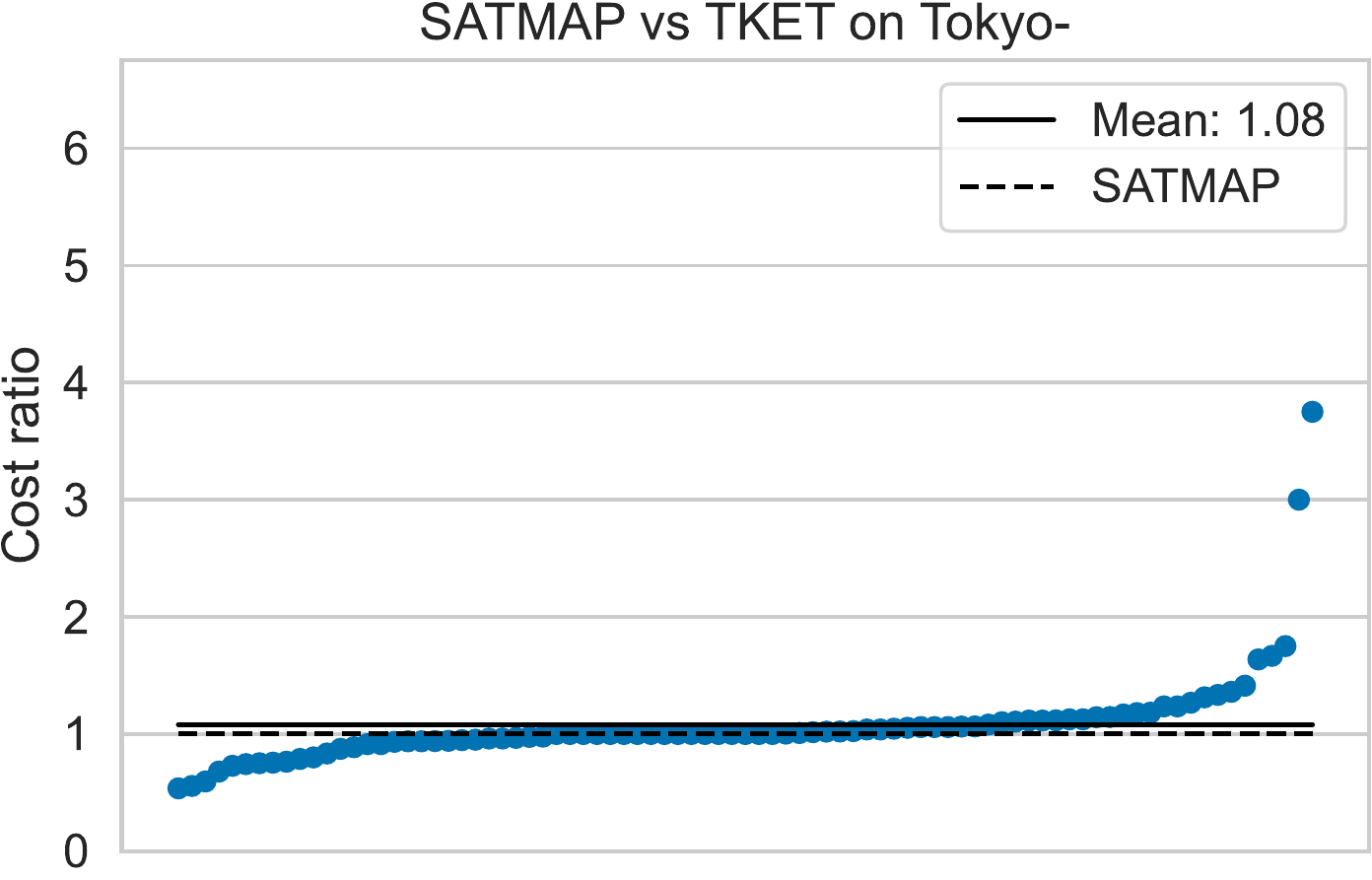}
  \end{subfigure}
  \caption{The cost of the solution produced by \tket divided by the cost of the solution produced by \ours on three different connectivity graphs: \archplus, \tokyo, and \archminus}
  \label{fig:rq4}
  \end{figure*}

\subsection*{Q4: Impact of Architecture}

\paragraph{Experimental setup} Finally, we investigated the effectiveness of \ours as compared to the best heuristic tool, \tket, when varying the properties of the connectivity graph.
We constructed two modified versions of the \textsc{ibm} Tokyo architecture: (1) increasing sparsity of the graph by removing diagonal edges (\archminus, \cref{fig:tokyom}), and (2) increasing connectivity by adding diagonal edges (\archplus, \cref{fig:tokyop}). The average degree of a vertex
in \tokyo is exactly halfway between \archplus and \archminus. We applied the same procedure as in the evaluation of Q2 with the use of these different architectures as the only difference.

\paragraph{Results}
In a similar manner to Q2, \cref{fig:rq4} shows the cost ratio on each benchmark for the three architectures.
On \archminus, heuristic tools and \ours produce very similar solutions.
The difference between the cost of the solution produced by \tket and \ours was less than 10 gates for 61 of the 85 benchmarks solved by \ours, with a mean cost ratio near 1. Results on \archplus are more in line with those on \tokyo, but with more variance across the benchmark set.
Again comparing \tket to \ours, the standard deviation in cost ratio on \archplus is 9.09 as opposed to 3.92 on \tokyo (excluding infinite ratios).  These results suggest the existence of two effects. First, heuristic solutions are well-suited to finding
near optimal solutions on sparse connectivity graphs. Second, the success of \ours on \tokyo as compared to \archminus and \archplus may indicate constraint-based tools are better suited to non-uniform architectures where the connectivity varies across qubits.
We observe the same behavior on \sabre and \mqth, so we focus on \tket here since it's the best-performing tool.

\textbf{Summary: heuristic-based tools are not robust to variations in the connectivity graph, tending to produce better results on sparse graphs (\archminus) than highly connected ones (\archplus).}

\subsection*{\diff{Q5: Scalability and Optimality}}

\paragraph{\diff{Time Limits}}
\diff{
First, we study the impact of the time bound on the number of benchmarks solved and the cost of the solution. 
We consider the following time limits (in seconds): 100, 300, 600, 1800, 3600, 5400, and 7200. 
We compare each time limit against the original 1800 seconds by computing the cost ratio like in Q2.
All other configurations are the same including the architecture, \tokyo.
As expected, the cost ratio decreases exponentially as the time allotted increases meaning solution quality improves with more time.
Additionally, the number of benchmarks solved and size of the largest circuit solved both increase given more time.
The change in the number of benchmarks solved across time is less dramatic, increasing from 103 to 111.}

\begin{figure}[t]
  \centering
  \includegraphics[width=\linewidth]{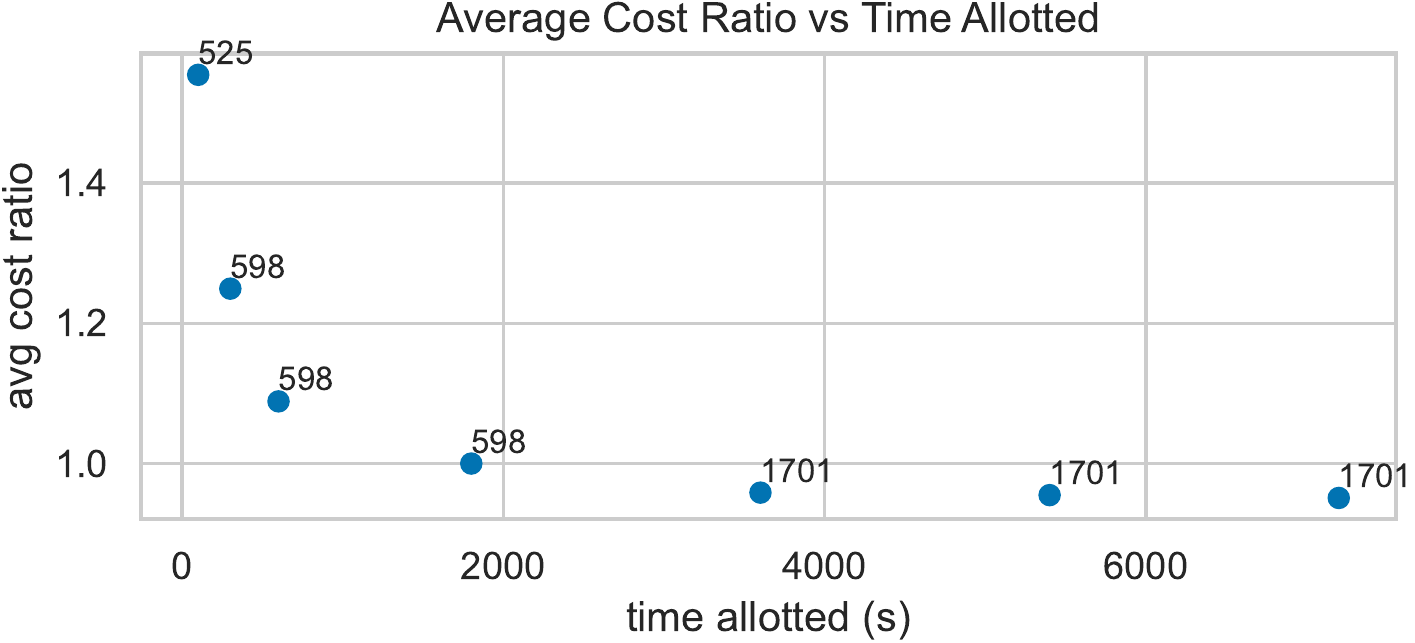}
  \caption{\diff{Comparison of average cost ratio across different time bounds with a baseline of 1800 seconds. Each point is labeled with the size of the largest circuit solved. }}
  \label{fig:cost_vs_timeout}
\end{figure}

\paragraph{\diff{Cost vs Circuit Size}}
\diff{
  Second, we analyze the optimality of the solution compared to circuit size.
  Since we do not have ground truth optimal cost, we use the cost ratio data from Q2 comparing \ours to
  the best performing heuristic tool, \tket.
  We observe a downward trend in cost ratio as circuit size increases, suggesting a loss
  in optimality as circuit size increases. 
  This is expected as the local relaxation creates more slices for larger circuits.
}

\begin{figure}[t]
  \centering
  \includegraphics[width=\linewidth]{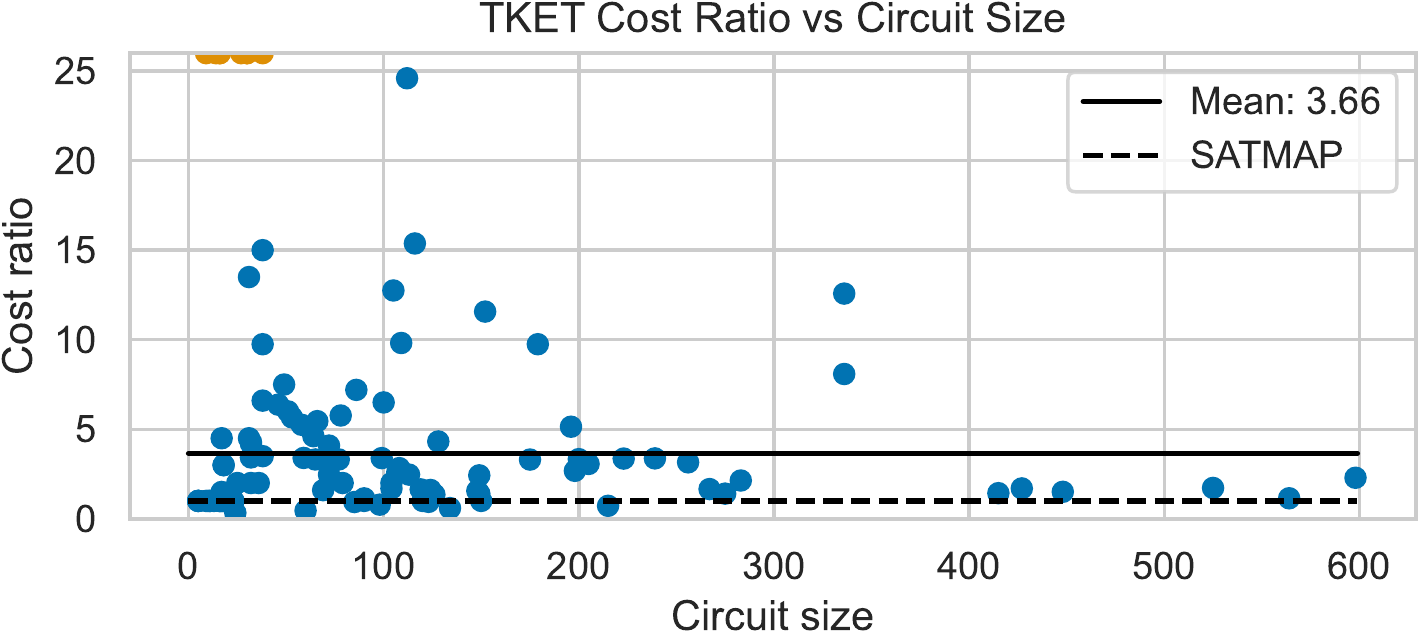}
  \caption{\diff{Solution quality compared to \tket across different circuit sizes.}}
  \label{fig:tket_cost_vs_size}
\end{figure}

\subsection*{\diff{Q6: Noise Models}}
\diff{
To demonstrate the versatility of our approach, we also use a \emph{weighted} \maxsat encoding to incorporate noise models, with the aim of maximizing the \emph{fidelity} of the output circuit. 
 
Weighted \maxsat is a generalization of \maxsat where soft constraints are assigned a positive weight, and the objective is to maximize the sum of the weights of the satisfied soft clauses. The \maxsat problem as defined previously is equivalent to weighted \maxsat where all weights are 1.
\begin{example}
  Consider the weighted \maxsat problem with one hard constraint and two soft constraints, 
  \begin{align*}
  \hard  &= \{ a \lor b \}\quad\\ \soft &= \{(\lnot a, weight: 5), (\lnot b, weight: 1)\}
  \end{align*}
The solution is $\model = [a \mapsto \false, b \mapsto \true]$, with a weight of 5.
  \end{example}

In the noise-aware version of \ours, we use soft-clause weights to encode the fidelity of operations, e.g, a  variable  $\vswap(\physq_1,\physq_2,k)$ is assigned a weight (probability between 0 and 1) corresponding to the fidelity of performing a \SWAP on the edge $(p_1,p_2)$. 
The result is an optimization objective equivalent to \olsq's.
We used the error rates supplied by the ``FakeTokyo'' backend from the \textsc{ibm} Qiskit development kit.

Fidelity maximization is a more complex objective than \SWAP minimization, so both tools, \ours and \olsq,
solved fewer benchmarks from the set of 160 within the same timeout from Q1. 
However, we observe an even bigger gap between the two tools, with \ours solving $\sim$4x more benchmarks: 
The fidelity 
maximization version of \olsq was able to solve 23 benchmarks, whereas \ours was able to 
solve 89. 
For the 23 benchmarks solved by \olsq, the 
fidelity achieved by 
by \ours was the same except for 5 examples, which incur a small fidelity reduction of 0.004 to 0.09 due to relaxation.
}

\section{Related Work}

\paragraph{Constraint-based Approaches}
Many prior works utilize constraint solvers for mapping and routing logical qubits. Several tools leverage \textsc{ilp} solvers to minimize the number of \SWAPs and circuit depth~\cite{bhattacharjee2019muqut,chakrabarti2011linear,bhattacharjee2017depth}. At the same time, several others translate \qmr to Boolean constraints and use \smt solvers~\cite{wille2019mapping,siraichi2018qubit}. Besides restricted connectivity, mapping logical qubits onto physical qubits can be challenging due to non-uniform gate latency and error rates. To that end, Tan et al. leverage \smt solvers to minimize the total circuit runtime~\cite{tan2020optimal} whereas Murali et al. use  
\smt solvers to maximize success probability by accounting for variability in gate errors~\cite{murali2019noise}. Furthermore, some tools cast \qmr as a path planning problem and leverage temporal planners~\cite{venturelli2017temporal,venturelli2018compiling}. However, most constraint-based solvers face severe scalability issues due to the exponential search space.

The closest work to ours is \olsq~\cite{tan2020optimal}
which uses a \emph{satisfiability modulo theories} (\smt) encoding
that is more efficient than earlier work~\cite{wille2019mapping}.
Our encoding is different in a number of ways: (1) We restrict ourselves to fully Boolean encoding, allowing us to sidestep the complexity of \smt vs \sat solving. 
(2) We model \SWAPs via Boolean variables in a view that mimics sketch-based program synthesis.
(3) We introduce novel relaxations that increase scalability while maintaining almost optimality.

\paragraph{Heuristic-based Approaches}
Due to the limited scalability of constraint-based tools, most industry compilers 
and open-source quantum compiler projects use heuristic methods for performing \qmr. 
For example, \textsc{ibm} Qiskit uses \sabre, a bidirectional local search algorithm that
slices the circuit into subcircuits and finds locally optimal mappings, similar to our locally optimal relaxation~\cite{li2019tackling}. 
In contrast, the \textsc{mqt} compiler uses a slow but exhaustive $\text{A}^*$ search~\cite{zulehner2018efficient}; the approach is made feasible by only applying $\text{A}^*$ between gates in consecutive topological layers of the circuit. Furthermore, recent work combines $\text{A}^*$ with novel search space pruning techniques to enable time optimal \qmr solutions~\cite{zhang2021time}.  
Tools like Enfield use subgraph isomorphism and token swapping~\cite{siraichi2019qubit}, 
whereas others use hierarchical product algorithms for modular architecture~\cite{childs2019circuit} 
and \SWAP networks~\cite{o2019generalized}. Majority of the heuristic methods try to minimize the distance between logical qubits and use greedy but efficient local search to reduce the number of gates and circuit depth.  For example, \textsc{tket} heuristic performs a greedy search to find an initial qubit mapping that results in the least number of \SWAPs and inserts routing gates by iteratively permuting the logical to physical mapping~\cite{sivarajah2020t,cowtan2019qubit}. Similarly, earlier work uses local permutations of physical to logical qubit map and sub-graph isomorphism to solve the \textsc{qmr} problem~\cite{maslov}. A large body of work focuses on restricted qubit architectures with 1D, and 2D nearest neighbor connectivity ~\cite{hirata2011efficient,saeedi2011synthesis,shafaei2013optimization,wille2016look}. Whereas recent works focus on \textsc{ibm} machines and develop greedy search strategies for \textsc{ibm} architectures~\cite{zulehner2019compiling, kole2019improved}. 
\diff{While others develop application-specific~\cite{guerreschi2018two, alam2020circuit} and noise-aware strategies that use hardware-level characteristics to improve fidelity~\cite{tannu2019not,das2021adapt,das2021jigsaw,jurcevic2021demonstration,tannu2019ensemble,patel2020veritas,stein2022eqc}.}

\section{Discussion} 

\paragraph{Importance of optimality}
Our results demonstrate a big gap in terms of added \SWAPs between heuristic-based \qmr algorithms and our constraint-based technique.
For near-term, noisy quantum computers, reducing the number of \SWAPs is critical for successfully executing quantum algorithms.
Therefore, our results indicate that, going forward, (1) we need to improve existing heuristic algorithms to bring them closer to optimal or (2)
improve the scalability of constraint-based techniques.

\paragraph{Scaling the \maxsat approach}
Our \maxsat approach produces significant improvements over existing optimal approaches.
However, \sat solving is an \textsc{np}-complete problem and scalability remains an issue.
We see two avenues for scaling our \maxsat approach to stay ahead of the growth in the number of qubits:
First, we can employ parallel \sat-solving strategies. All of our experiments used a single-threaded solver. Today, there are \sat solvers that can run on cloud infrastructure with impressive improvements.\footnote{\url{https://satcompetition.github.io/2021/}}
Second, at the algorithmic level, we can combine our \maxsat approach with
heuristic approaches. For instance, we can only solve the mapping constraints (optimally)
and leave the routing process for a heuristic approach or an approximation algorithm~\cite{siraichi2018qubit}.

\paragraph{Future architectures}
Quantum computing is a field in flux, and there is no clear indication of how future connectivity graphs will look like, as it depends on the underlying physical substrate used and engineering advances.
The ideal, of course, is to build a device with as much connectivity  and as little cross-talk as possible.
Our results demonstrate that for higher connectivity, the performance of heuristic approaches diverges from the optimal. 
So it is prudent to robustify heuristic approaches to changes in the connectivity graph to accommodate the rapid developments \nisq computing hardware.
\section*{Acknowledgments}
We thank the anonymous reviewers for their thoughtful feedback and Rui Huang for providing scripts to generate \qaoa circuits. 
This work is supported by \textsc{nsf} grants \#1652140 and \#2212232 and awards from Meta and Amazon. This research is also partially supported by the Vice Chancellor Office for Research and Graduate Education at the University of Wisconsin–Madison with funding from the Wisconsin Alumni Research Foundation.


\bibliographystyle{IEEEtran}
\bibliography{IEEEabrv, refs}

\end{document}